# Mesoscopic collective dynamics in liquids and the Dual Model

## *Fabio Peluso*

*Leonardo SpA – Electronics Division – Defense Systems LoB*

*Via Monterusciello 75, 80078 Pozzuoli (NA) – Italy*

mailto: fpeluso65@gmail.com.

## Abstract

A microscopic vision is presented of a Dual Model of Liquids (DML) starting from a solid picture. The task is accomplished firstly by showing how a series of experimental evidences and theoretical developments on liquid modelling, gathered for the first time, can be framed in a mesoscopic view of liquids, hypothesized as constituted by a population of dynamic aggregates of molecules, diving in an ocean of amorphous liquid. The pseudo-crystals interact with the rest of the liquid through harmonic elastic waves and anharmonic wave-packets propagating within and among the structures. The anharmonic interaction term is derived from "first principles"; it allows the exchange of energy and momentum between the wave packets and the molecule's clusters, determining the displacement of the latter within the medium, and the redistribution of the energy between external Degrees of Freedom (DoF) and internal collective degrees of the clusters. Among the novelties of this model is that it provides quantitative expressions of various extensive thermophysical properties. The introduction of the statistical number of excited DoF allows to bypass the problem of other dual models which are sometimes unable to correctly reproduce the expressions for those thermophysical quantities showing deviations due to the activation/deactivation of internal DoF. The interpretation of the relaxation times is given, their Order-of-Magnitude (OoM) calculated and the way in which these times are involved in the different phases of the collective dynamics of liquids discussed.

A comparison is provided with results obtained in the frame of Phonon theory of Liquid Thermodynamics, as well with the forecasts for the viscoelastic transition regions and with systems exhibiting k-gap.

In the last part of the paper, theoretical insights and experiments are suggested as potential directions for future researches and developments.

## 1. Introduction

Liquid and solid states are thermodynamic concepts correlated with the relative motions of the constituent molecules and the rates at which energy and momentum are redistributed among the available DoF. The liquid state of matter is that which exists in the narrowest temperature range compared to all the others. Since it is the most important one underlying the phenomena that regulate and influence the development and persistence of life on our Planet (together with the gaseous one), many efforts have been and are dedicated by scientists to identify a unifying physical model for the liquid state.

Gas, liquids and solids are characterized by very different behaviour and that of liquid state is sometimes contradictory. Just to cite some, on macroscopic scale the molecules of solids are strongly bounded to their positions for long times, while liquids, as gases, besides being incapable of opposing to tangential stresses, are also characterized by diffusive phenomena evolving on short timescales. Melting is accompanied by a relatively small amount of volume, while the liquid–to-gas transition is characterized by a huge increase of specific volume. Consequently intermolecular distances in solids and liquids are quite the same and the latent heat of fusion is much smaller than the latent heat of vaporization. These facts allow us to affirm that binding energies are similar in solids and liquids, while in liquids long-range order is reduced, making a liquid isotropic, or amorphous, like a gas.

On a macroscopic scale liquids are being studied with the methods of fluid dynamics, while on a microscopic scale one resorts to those of quantum mechanics. Already in XIX century it was understood that fluid dynamics was sometimes unable to explain the behaviour of liquids, in particular those considered anomalous when compared to what expected for a condensed gas, as they were initially classified according to their macroscopic behaviour. The advent of quantum mechanics then paved the way towards liquid models seen as "molten solids", allowing scientists to realize that it became necessary to investigate an intermediate, mesoscopic scale for liquids. This intermediate scale, where liquids deviate from their macroscopic behaviour, although the dynamics still involve a large number of molecules, was the object of theoretical investigations originally proposed by Debye [1,2], Brillouin [3,4], and Frenkel [5]. The peculiarity of the liquid state makes its modelling very difficult, because it is not possible to benefit from the simplifications that are adopted for gases and solids. In fact, in gases, since the molecules are very distant from each other, the interactions are treated as perturbations of the trajectory of the molecules themselves. In solids, because the atoms are held in their positions by very intense forces, it is the displacements from the equilibrium positions that are very small, and therefore susceptible to simplifications in the modelling process. Landau condensed the above concept as *liquids do not have small parameters* [6]: the molecules interact by means of intense forces, comparable to those in solids, and their displacements are very large, comparable to those of the molecules of a gas. The relaxation mechanisms become fundamental in the dynamics of molecules in liquids and in the processes of redistribution of energy and momentum among the available DoF. Due to the complexity of liquid behaviour on different scales, a general and unifying physical modelling of liquid structure is still an open question of modern physics.

These facts represent the main obstacles that physicists have to face in modelling liquids. Various lines of thought have been developed over the past 150 years, recently reviewed in critical way by Chen [7], both with elaborations on different models developed for several extensive liquid quantities, such as the specific heat, thermal conductivity, and viscosity. The review shows, as the author himself underlines, that "current understanding of thermophysical properties of liquids is still poor and theoretical tools to study them are not well developed yet." One line of thought tries to understand transport in liquids by modifying the Boltzmann theory of gas, as Enskog did (see ref. [9] in [7]). Another line, pioneered by Einstein, ([see ref. [10] in [7]) consisted in determining the molecular size by studying the Brownian motion; this approach evolved in two directions: (a) reduction of the Liouville equation describing the N-particle distribution function in the system, and (b) the linear response theory based on the perturbation solution of the Liouville equation for the time-evolution of the N-particle distribution function. The former forms the basis of hydrodynamic treatment of liquids, and the latter is the foundation of molecular dynamics simulations, a major tool currently used in studying liquids. The third line of thought consists in starting from the solid picture, it arguably can be dated back to Maxwell [8], successively reworked by Debye, Brillouin and Frenkel [1- 5].

Many studies and books have been devoted to understanding liquids and significant progress has been made ([8,9], see also Refs. [8,9,15,23-29,31-38] of [7]). Despite that, our current ability to predict the thermodynamic and transport properties of liquids is still immature. Chen [7] provides a very detailed picture of the development of liquid state models, roughly grouped into (1) modelling the partition function, (2) statistical formulation, (3) numerical simulation, and (4) phonon theories. Because of the microscopic characteristics of DML, we will limit here to recalling the most pertinent bibliographic sources (many of them not mentioned in Chen's work); however, the content of this manuscript does not in any way want to be in contrast with the most widely diffused models of liquid state, but simply be an alternative point of view of a liquid physical model that, by its nature, cannot be checked and verified directly but only through indirect measurements, very often subject to different interpretation, and as such being the perfect example of the Popperian non falsifiability.

The paper is organized as follows. In section 2 after a brief historical overview of the experimental evidences of collective dynamics in liquids, these are compared with the results obtained in the recent scattering experiments on liquids. In particular, the role of the relaxation time is discussed. In section 3 the DML is presented through the description of the thermal, or elastic, excitations, and of their interaction with the dynamic icebergs: in paragraph 3.2 it is shown how the expression of the interaction term can be deduced on the basis of "first principles" of wave mechanics; in paragraph 3.3 the dynamics of the interactions between wave packets and *liquid particles* is analyzed, and the expressions of thermal capacity, thermal conductivity, and in general of the contribution to the internal energy of the liquid due to the presence of wave packets, are derived; finally, section 3.4 is dedicated to the step-by-step description of the interaction process and to identify the several relaxation times that characterize it, to calculate their OoM and finally to compare the results with the forecasts of the viscoelastic cross-over in liquids. Section 4 is dedicated to the numerical simulations of the Dual model of liquid state, in particular the most recent results of the calculation of the internal energy of the liquid and of its specific heat are discussed by comparing various theoretical models with experimental data. In section 5 the results obtained in paragraph 3 are discussed in detail and compared with the recent theoretical and experimental achievements on the structure of liquids at the mesoscopic level. The role of interaction and relaxation times is deeply analysed, as well the k-gap for transverse phonons. Finally, in the "Conclusions" further theoretical developments, numerical modelling and experiments are proposed, among other things, to be conducted also in the virtual absence of gravity to reveal and study the dynamic evolution of wave packets propagation in liquids.

## 2. Brief historical overview and recent experimental evidences of collective dynamics in liquids

The hypothesis that thermodynamic equilibrium properties of liquids could be interpreted in terms of those of thermal excitations, supposed as consisting of high-frequency wave-packets, was first introduced by Debye, subsequently reworked by Brillouin, Frenkel and Lucas [10]. Debye extended to liquids his theory of heat capacity of solids, neglecting transverse waves since liquids lack of rigidity. He assumed the longitudinal high-frequency waves to be very similar to the collective oscillations of solids. Thermal energy in a liquid is then equally distributed among the various DoF of the molecules and the phonons. Brillouin accepted the argument of Debye but observed that, neglecting transverse waves, the Debye model would have lead to the paradox of a higher limiting Debye frequency $\nu_D$ in liquids than in solids. A fundamental forecast of Brillouin, later reworked by Landau [11] and Fabelinskii [12], was that a light beam crossing a liquid would have been affected by a Doppler shift. The cause for this phenomenon is that a density fluctuation is induced in the system under investigation by means of a light scattering experiment by the impinging probe. This (neutron or X-ray) is scattered by the target particle, giving rise to the spectroscopic pattern proportional to the dynamic structure factor, $S_{EQ}(k,\omega)$ [13-16], which is a function of the energy and momentum exchanged in the scattering event. This was confirmed for the first time by Meyer and Ramm and by Raghavendra-Rao, whose experiments indeed showed that light diffusion in liquids is accompanied by a frequency shift, with the emergence of a Lorentian triplet: a central, unshifted line (Rayleigh), accounting for elastic scattering, and two side-lines (Brillouin "Stokes" and "anti-Stokes" doublet) symmetrically shifted by

1. $\Delta\omega_B(k) = \pm c_s q$

with respect to the central line [13]. In Eq.(1) $c_s$ is the sound velocity in the liquid, $q$ the wave vector. As is well known, it is believed that these components of the spectrum of a liquid are due to the inelastic collision between an impinging photon and a phonon, or sound mode, of the fluid in which the photon gains or loses energy. The HWHM of these two lines give the lifetime of a classical phonon of wave vector $q$, while that of the unshifted line gives the thermal diffusion coefficient of the liquid, $D_T = K/\rho C_P$. When the three lines are well resolved, a light scattering measurement provides experimental evaluation of relevant liquid parameters.

Another important key point raised by Brillouin is the value of the specific heat of monatomic liquids, $C_V$. From a classical point of view, $C_V$ should increase with temperature from the melting point to the critical point; measurements show instead that it decreases from about $3\mathcal{R}$ at the melting point down to $2\mathcal{R}$ at the critical point, where $\mathcal{R}$ is the gas constant [17]. The interpretation of this behaviour in the frame of the Debye model was provided by Brillouin [4] and Frenkel [5], by admitting that a pseudo-crystalline structure still survives in liquids above the melting point. This point will be deepened in §4 and in the *Discussion*.

Around mid 60's the study of the static structure factor $S(q)$ obtained by means of INS and IXS showed that in liquids, and especially in water, $S(q)$ was characterized by the presence of elastic contributions that, in turn, contributed an enhancement of the $S(q)$ profiles at high frequencies. Sköld [18] interpreted $S(q)$ as a measure of the effective number of atoms contributing to the scattered intensity at a given exchanged wave vector $q$. Years later [19-20] the advent of new techniques allowed to perform IXS scattering in the THz regions, this argument became fundamental in ascribing the $q$ dependence of the scattered spectra to the higher inertia of the target system due to the large number of atoms participating to the collective response of the target system. In particular, the first neighbour movements become more correlated when $q$ matches the inverse of the first neighbour's separations and $S(q)$ approaches its maximum [20]. The forerunner and most surprising experimental datum on the dynamics of liquids at high frequencies is that found by Ruocco and Sette [21]. In 1996 they measured by IXS experiments in liquid water at ambient conditions the propagation speed of elastic waves, and found it equal to $3200\,m/s$, i.e. more that double that at traditional frequencies, $1500\,m/s$, and very close to that of ice $I_h$, $4000\,m/s$. Starting from the experimental results obtained in various liquids with IXS techniques, [19-35], the evidence that ephemeral pseudo-crystalline structures "*exist and persist*" in liquids has gradually consolidated [7, 36-37], their size and number depending on the liquid temperature (and pressure). This picture resembles in some way that currently accepted for water, that is believed to consist of hydrogen-bonded clusters in dynamic equilibrium [7]. It is worth noticing, among others, the pioneering experimental investigations by Grimsditch [36] and those by Giordano and Monaco [37], who performed first measurements across the glass-liquid and solid-liquid transitions, respectively. Giordano and Monaco report on a comparison of the collective excitations in liquid and polycrystalline sodium. As it concerns liquid sodium, it exhibits acoustic excitations of both longitudinal and transverse polarization at frequencies strictly related to those of the corresponding crystal. The relevant difference between the liquid and the crystal is the broadening of the excitations in the case of liquid because of an additional disorder-induced contribution coming into play. The Authors deduce a direct connection between structural and dynamic properties of liquids, with short range order and overall structural disorder characterised by specific fingerprints.

These structures can be discovered only when high frequency measurements are carried out, typical of INS and IXS techniques, the wavelengths of the radiation involved being small enough to reveal their presence. In the DML framework these structures interact with the rest of the (amorphous) liquid by means of anharmonic interactions which arise at their border, a quasi-elastic propagation within the clusters is transmitted to the amorphous matrix in the form of anharmonic wave-packets, and vice-versa. This idea is rendered in Figure 1. Unlike in crystalline solids, the anharmonicity allows a non-infinitesimal duration of the interactions, with consequent transport not only of energy but also of momentum, and furthermore exchange of energy with the internal DoF of the solid-like dynamic structures. Since this dynamics occurs mainly at high frequencies, and involves only the DoF of the lattice, we introduce the parameter $m$ (see Eq. (15)) to account for the average number of DoF participating to the collective dynamics, This model provides an intuitive way of describing liquids how they emerge from the experiments. On the other hand, it would not be surprising if the transition from the solid to the liquid state does not suddenly occur in the narrow range of the triple point, but rather over a much wider range of temperature and pressure, this fact being the explanation of the "*existence and persistence*" of pseudo-crystalline structures in the liquids above the melting point.

Frenkel and Brillouin had not gone so far as to model the interaction mechanism and to write the related interaction term between elastic waves and molecule clusters, as well as to provide the basis for writing the thermodynamic and fluid dynamic coefficients. Frenkel himself admits, in the introduction of his book [5], that "*the kinetic theory of the liquids* (he proposes), *even being in a very crude and incomplete form, may serve to attract the attention of other scientists to this subject and accelerate its further development*." Recent experimental findings however [19-35] have given credit again to their early intuitions. The ideas of Debye, Brillouin and Frenkel were not considered for about 40 years, until Fiks used them again in the early '60s to explain thermal diffusion in liquids [38]. Fiks derived an expression for the coefficient of thermal diffusion, i.e. with a temperature gradient externally applied to the system. Subsequently Fiks used the expressions obtained to describe the electronic drag and thermal diffusion in metals [39]. In a very interesting paper Andreev [40] assumes that a liquid consists of two weakly coupled systems, the phonons and the remainder of the liquid. He calculates the effects connected with the presence of weakly damped phonons in a normal liquid. In particular, he provides the expressions for mechano-thermal and thermo-mechanical effects, as well as the propagation of shear oscillations. Similar ideas were later developed and extensively applied by Gaeta and co-workers to describe the behavior of non-isothermal liquids, allowing to interpret some of their typical phenomena, such as thermo-osmosis, thermodialysis, the heats of transport and the Soret effect [41-45]. In particular the model provided the heat diffusion direction of the solute, or of the solvent, with respect to that of the applied temperature gradient.

All the works described above have as common leitmotiv the presence of a temperature (or concentration) gradient externally applied to a liquid solution. In the present paper this limiting constraint is completely removed and overtaken; it is built an interaction model between wave-packets and the clusters of liquid molecules of a pure liquid, regardless of whether external gradients are applied to the system. The cue for this insight comes from the experimental evidence cited above of the presence and permanence of ephemeral pseudo-crystalline dynamic structures in isothermal liquids. Among the main results is that DML allows to identify and quantify, at least the OoM of the most relevant relaxation times which characterize the dynamics of the interaction process occurring in the liquid at mesoscopic scale, between the "*liquid particles*", (the clusters of molecules, in the general meaning of a dynamic aggregate of molecules) and the "*lattice particles*", i.e. the elastic wave-packets. As we will see later, a peculiarity of the interaction term between *lattice particles* and *liquid particles* is its anharmonic character, that allows the exchange not only of energy but also of momentum between wave packets and clusters, thus determining both the displacement of the latter within the medium, and the redistribution of the energy between external, or translatory DoF of the clusters, and internal collective, vibratory DoF.

There were no relevant advancements in the development of such a model of liquid state until the end of the 70's – beginning of the 80's (apart the contribution of Fiks and Andreev cited before) when many papers begun to be published on the study of hydrodynamic fluctuations in liquids out of equilibrium. In particular, some Authors [46-52] calculated the form of the light spectrum scattered from a liquid near or far from thermal equilibrium. According to [46-50], in case of small gradients, by using perturbation theory around a state of equilibrium, the line shape and intensity are modified with respect to the ordinary Landau-Placzek formula [11-12]. In particular, the central Rayleigh line remains the same as in equilibrium, while shape and intensity of the two Brillouin lines deviate from their equilibrium values by terms proportional to the temperature gradient. The dynamical structure factor for a fluid in non-equilibrium steady state (NESS) with a small temperature gradient (SG) is given by:

2. $\quad S_{NESS}^{SG}(q,\omega) = S_{EQ}^{R} + \sum_{\sigma=\pm 1} S_{EQ}^{B}(\sigma) \left[ 1 - \frac{\sigma y_\sigma c_s \hat{q} \cdot \nabla T}{\Gamma_s q^2 T} \right]$

where $\hat{q}$ is the unitary wave-vector, $\sigma = \pm 1$ according to the Brillouin peaks, $\Gamma_s = (4\nu/3 + \zeta/\rho) + (\gamma - 1)D_T$ the sound-damping constant (the product $\Gamma_s \cdot q^2$ is the line-width) and $y_\sigma = 1 + \frac{[(\Gamma_s q^2/2)^2 - (\omega - \sigma c_s q)^2]}{[(\Gamma_s q^2/2)^2 + (\omega - \sigma c_s q)^2]}$. As for the intensity of the light scattered by the sample, although the total intensity remains unchanged, a dependence on the wave-vector $\hat{q}$ appears in the two Brillouin lines:

3. $$I_{NESS}^{SG} = I_{EQ}^R + \sum_{\sigma=\pm 1} I_{EQ}^B \left(1 - \frac{\sigma c_s \hat{q} \cdot \nabla T}{\Gamma_s q^2 T}\right).$$

The analysis of Eqs. (2) and (3) shows that i) the central line is not affected by the temperature gradient and, ii) although the positions of the Brillouin lines are unchanged, they are asymmetric and non-Lorentzian, the asymmetry factor being proportional to the ratio $\nabla T/q^2$. The same proportionality is found in the difference of intensity between the two Brillouin lines. According to [46-51], the $q^{-2}$ dependence in both shape and intensity indicates the presence of long-range correlations in the fluid out of thermal equilibrium, proportional to the inverse of the average intermolecular distance. Besides, the asymmetry in the Brillouin lines is due to a coupling between two sound modes with the heat current produced by the temperature gradient. Owing to this coupling, the compression waves travelling in the same direction as the macroscopic heat flow will have a larger amplitude than waves travelling in opposite direction. We will return on this aspect in §3 and in the *Discussion*.

When large temperature gradients are applied to the liquid, shape and intensity also of the central Rayleigh line show deviations from equilibrium values, proportional to the square of the temperature gradient [50]:, the intensity is enhanced when compared to its equilibrium value, while the deviation of intensities of the Brillouin lines from their equilibrium values has a rather complex dependence upon the temperature gradient.

The interest in calculating the non-equilibrium correlation functions was stimulated by the fact that, owing to the presence of the gradients in the system, such correlation functions have a much longer range in space than their equilibrium counterparts, and that these long-range non-equilibrium correlations could be detected by means of light-scattering techniques. The above theoretical works were motivated by the results found by Beysens and co-workers [53] who performed very-low-angle Brillouin light-scattering experiments in liquid water subjected to a temperature gradient. Their results showed the Brillouin asymmetry in shape and intensity foreseen by Kirkpatrick [49-50]. Subsequently, other very-low-angle light-scattering experiments were performed with various liquids far from critical conditions [54-56]. Experimental results confirmed the theoretical forecasts obtained with the non-equilibrium fluctuations approach, and the observed phenomenology of asymmetry in the Brillouin lines was interpreted as due to an asymmetry in the phonon flux induced by the temperature gradient [53-56]. Because Brillouin doublet accounts for energy and momentum transferred in inelastic scattering events, and the central Rayleigh line accounts for elastic scattering, at large temperature gradients, while the correlation lengths still increase with respect to the isothermal or near to thermal equilibrium values [49-50], it is expected that inelastic contributions of *phonon ↔ liquid particle* interactions also have a large relative importance in the overall dynamic processes.

So far the early theoretical forecasts and first experimental evidences of the presence of long-range correlations in liquids. Very interesting results came also from the numerical modelling of isothermal (equilibrium) liquids. In the seventies Rahman and Stillinger [57] set-up a numerical model to calculate $S_{EQ}(q,\omega)$ for liquid water based on the Molecular Dynamics approach. Their model analysed the fluctuation phenomena occurring in a system of 216 water molecules kept at ambient steady conditions. Two interesting results were obtained, namely that transverse currents were present in the liquid in the form of propagating collective modes, and that the spectrum of density of fluctuations exhibited a secondary maximum at a much higher frequency than the usual sound propagation frequency . Due to the difficulties of resolving the spectral lines in liquids at ordinary light frequencies, experimental verifications of the numerical simulation of Rahman and Stillinger came only in the mid of 80's thanks to the INS technique [58]. The study of the Brillouin doublet generated in IXS experiments confirmed years later the previous experimental results with a better resolution. Experiments confirmed that in isothermal liquids there are two branches of collective modes, one is the known non dispersing mode propagating at the ordinary speed of sound, the other is linearly dispersing with the frequency and propagates at a speed higher that that of ordinary sound. In the case of water the speed of the second branch was detected at $\approx 3200 m/s$ [23-35,59], and the cross-over in between the two speeds. This result, already foreseen in [57] and later confirmed in other works [29], was correlated with the H-bond arrangement typical of liquid water at mesoscopic scale, marking the transition from ordinary to "fast" sound. The study of this transition as function of frequency and temperature allowed to relate this phenomenon to a structural relaxation process and showed many analogies with glass-forming systems [22-37].

The occurrence of a solid-like value of sound propagation in liquids can be theoretically explained following the points of view of Brillouin and Frenkel. Indeed we may distinguish two limit regimes for propagation of perturbations in liquids. One is the "normal" hydrodynamic regime, in which one considers the medium as a continuum and the excitations on a so long timescale that the system may be assumed to be in thermodynamic equilibrium. This regime is also usually referred to as *viscous*. The second is the "solid-like" or *elastic* regime, in which the dynamics becomes that of a free particle between successive elastic collisions. The investigation of collective dynamics becomes of particular interest when intermediate time- and length-scale are considered, i.e. distances comparable to those characterizing the structural correlations among particles, and times comparable to the lifetimes of such correlations.

Although these two extreme behaviours, viscous and elastic, are well known in physics, the intermediate situation is still well far from being fully characterized. This intermediate range is referred to as *viscoelastic* regime.

Experimental results obtained in the last forty years allow to demonstrate the validity and limits of the approach adopted in hydrodynamic fluctuation modelling (many references may be found in [60]), and to advance new insights on the mesoscopic structure of liquids [19-20,22-24,61]. If at low frequencies, i.e. large wavelengths, it is not possible to "see" the mesoscopic structure of liquids because they oscillate under the effect of pressure waves, travelling at the speed of sound, we will show that their behaviour at very high frequencies, or very short wavelengths, can be explained by means of a liquid mesoscopic structure organized on dynamic solid-like clusters, also called elsewhere solid-like or pseudo-crystalline structures [19-35,59,61-72] (named here also "dynamic icebergs"). Such clusters are indeed dynamic objects in continuous rearrangement, in particular in water their ephemeral presence is enhanced by the characteristics of H-bond, as pointed out also by Chen.

The experimental value of $\approx 3200 m/s$ for the speed of sound in liquid water [21,25-26,31] may be explained by assuming that water (and liquids in general) behaves as a rigid network of molecules, i.e. intermolecular forces and molecule arrangement are the same as in ice on the length scale and during the time lapse of the relaxation time. Indeed both the energy and the speed of the acoustic wave generated after the collision of the probe with the target particle of the medium increase from the viscous to the elastic regime, i.e. upon frequency increase, (positive dispersion vs frequency, PSD [19]), while the opposite is true for the viscosity. Consequently, the trend of the liquid physical parameters with frequency represent fingerprints of relaxation phenomena typical of collective dynamics. This conclusion is the same that Brillouin and Frenkel had reached in their pioneering works [4,5], as demonstration of their ingenious intuitions which anticipated what has been experimentally assessed only a century later. Frenkel in particular introduced the idea that liquids are constituted by a very large number of randomly oriented dynamic (ephemeral) pseudo-crystals of submicroscopic size[1], that continuously rearrange their composition in terms of atoms or molecules. To explain this concept, he used the relaxation time $\tau_F$, introduced for the first time by Maxwell [8], as the average time between particle jumps at one point in space in a liquid [5]; its inverse, $\nu_F = 1/\tau_F$, is the frequency of occurrence of the particle jumps. The presence of relaxation time distinguishes liquids from gases; in the latter every collision is independent from any other, and every collision is equal to the previous and successive ones. The heat motion in liquids close to the crystallization point has the same character as in solids, consisting of oscillations of molecules around their equilibrium positions. The positions of atoms in liquids are of course not permanent but temporary. After performing a number of oscillations around a given position, the atom can jump to another equilibrium position, far $\delta$ from the previous, of the same order of magnitude as the average distance among molecules in that liquid. This step-by-step wandering, shown in Figure 2, lasts $\tau_F$, is a sort of self diffusion motion, leading to a gradual mixing up of all the atoms [11,29]. It shall have a simpler character in liquids because of the absence of definite lattice sites. $\tau_F$ has two intrinsic physical limits [24,62-63,65]: the upper, at low temperatures, is located just above the crystallization point; the lower, at high temperature, corresponds to the minimum value of the Debye vibration period, $\tau_F = \tau_D = 1/\nu_D \approx 0.1 ps$. In the upper limit two successive jumps occur every , $\tau_F \approx 10^2 \div 10^3 s$ typical of the glass transition, while in the lower limit the relaxation time becomes comparable with the shortest vibration period. Thus liquids may be distinguished from solids or gases by means of the values of $\tau_F$. If, at a given temperature, $\tau_F$ is much longer than the characteristic time $t$ of a perturbation (or $\nu_F$ much shorter than the frequency $f$), the medium is seen by the perturbation as a solid and the particles have no time to rearrange. The emerging spectrum is that of a solid [35], while the speed of sound is much higher. This case is that of "fast sound", in which sound velocity is very close to that of the corresponding solid phase, a situation typical of the viscoelastic regime, where viscous and solid-like elastic properties are combined, as early suggested by Maxwell and Frenkel.

---

[1] In his book [5] Frenkel actually affirms that this idea was introduced by Stewart around 30's; Stewart proposed to denote such submicroscopic crystals, consisting of few tens of molecules, at most, by the term "cybotactic groups" (or regions), and assumed them to be connected with each other by thin layers of the wholly amorphous phase.

The first to assert [64] that *"the distinction between liquids and solids is quantitative and not qualitative"* was Eckart in his papers on the anelastic fluid [73-74], the border line being the relaxation time $\tau_F$, or the frequency $\nu_F$. Eckart got in his papers many other interesting results, as that the dependence on frequency becomes the propagation velocity of isentropic longitudinal waves a complex number. The problem of the relaxation times occurring in the dissipative processes was faced by neglecting all the possible causes of dissipation except the collisions (as in the DML), which give origin to the viscosity.

Liquids do not support all vibration modes as solids, but only those above $\nu_F$. More precisely, liquids support 1 longitudinal and 2 transversal modes at low temperature; as far as the temperature increases the capability of supporting shear stresses is lost, and only the longitudinal mode survives. This limit provides the border line between vibratory, or oscillatory, motions and purely diffusive motions.

Further implications and results of the DML are illustrated in what follows. In a series of recent papers [61-65,67-72] Trachenko and co-workers proposed the Phonon theory of Liquid Thermodynamics (PLT). In a similar frame as that of DML, PLT considers the liquid as made of two interacting subsystems [68]; the Authors start from a solid-like expression of the Hamiltonian in harmonic approximation, i.e. without interaction, that they extend to consider anharmonic contributions by introducing an appropriate interacting term [61,71]. This approach, that recalls Eyring's theory [75], has many interesting results. First, it uses a unique Hamiltonian for the solid, liquid and gas states of matter. Second, it provides for liquids two distinct solutions, depending on the value of the frequency: for frequencies lower than the Frenkel frequency, $\omega < \omega_F$, the system is characterized by one longitudinal mode, while for frequencies higher than $\omega_F$, the system exhibits three modes, namely two transverse non massive and one longitudinal massive. The theoretical expression for the specific heat $C_V$ obtained in the PLT has been validated in 21 different liquids [62], covering the cases of the solid, glassy, liquid, gas and quantum liquids states of matter, giving for $C_V$ a theoretical limit value of $3\mathcal{R}$ for solids down to $2\mathcal{R}$ for liquids [61,64,71]. DML too provides an expression for $C_V$ as function of the number of excited collective DoF. We will calculate $C_V$ in Section 3.3. In a separate paper [76] the dependence on temperature of $C_V$ is analysed and it is shown that the theoretical expression obtained in the DML is in line with the experimental results; besides, its comparison with the analogous one obtained in the PLT allows to get interesting insights about the number of collective degrees of freedom available in a liquid and on the value of the isobaric thermal expansion coefficient. A limitation of the PLT is that this model only deals with intermolecular modes and neglect the intramolecular ones; DML tries to cover this gap with the aid of the parameter $m$

It is important to highlight that the DML proposed here is not in contrast with the PLT, on the contrary they face the problem of liquid model from two different and complementary approaches, as more clearly specified in the *Discussion*. PLT and DML consider liquids as dual systems, PLT faces the problem from a thermodynamic statistical point of view, while DML faces the microscopic vision of the Dual System introducing an elementary interaction involving the two sub-systems, wave-packets and *liquid particles*. DML allows to determine the contribution to some fundamental parameters characterizing the liquid medium due to these interactions, such as thermal conductivity, specific heat, diffusion coefficient, molecule drift velocity, etc. An apparently simplifying hypothesis is that the collisions between wave packets and pseudo-crystalline structures have a single-particle character, as in the Boltzmann-Maxwell statistics. This hypothesis is however supported by the wave packet' density estimation (§3.4), which appears to be of the same OoM as the density of molecules, if not lower (see *Discussion*). Incidentally, it is very relevant the opinion of Chen [7] about the PLT dual model, or in general about the models based on the Frenkel picture, as the more fruitful theory when compared with others.

Before going on to describe the DML, it is worth pointing out right away that it does not want to be in contrast with those models derived from the continuum mechanics but, starting from the experiments, it provides a possible alternative vision of how liquids are organized at microscopic (or rather mesoscopic) scale. The model therefore has no claim to rewrite pages of physics that are perfectly framed by continuum mechanics (a summary of the different approaches to liquid modelling may be found in the Chen' review [7]). The classical hydrodynamic theory for continuous media consistently describes the dynamics of density fluctuations in simple liquids at quasi-macroscopic distances and timescales by means of the Dynamic Structure Factor, $S(k,\omega)$. The evolution of the $S(k,\omega)$ shape across various dynamic regions is characterized by the average inter-atomic distance $d$ and inter-collision time $\tau_{coll}$, that are usefully compared with the probed distance, $l = 2\pi/k$ and time window, $t = 2\pi/\omega$. From experiments one may consider two distinct situations, namely [19-20,24]: a) $l >> d$ and $t >> \tau_{coll}$ and b) $l << d$ and $t << \tau_{coll}$. Being in a) the system a continuous and isotropic medium, its dynamic response is the average over a large number of interactions. Within this limit, any information on the microscopic structure and internal DoF of the sample is lost, and the hydrodynamic theory for continuous and homogeneous media consistently accounts for the dynamic response. Case b) represents the opposite limit and the event under investigation becomes the collision between a single atom and the probe particle (e.g. a photon or a neutron). In this case the shape of the spectrum is nothing else than the momentum distribution of the struck particle in its initial state.

Although the $S(k,\omega)$ shape is well-known in these two limits, a theory predicting its evolution in between them has not been set yet up. To cover the gap, the so-called Generalized Hydrodynamics (GH) theory of density fluctuations has been developed. Because along the crossover between the two limits density fluctuations are strongly coupled with molecular DoF, the interpretation of $S(k,\omega)$ becomes particularly complex. Owing to the lack of a theory predicting such a Q-dependence, the various theoretical attempts to extend the hydrodynamic description mostly require phenomenological recipes. In particular the GH models rely on the assumption that the deviation from the Rayleigh–Brillouin shape is smooth enough to leave the formal structure of the hydrodynamic spectrum unaltered. The Q-dependence is determined introducing some free parameters, whose amplitude is determined *a posteriori* by best fitting the GH model to the line-shapes measured (or computed) at various Q values. The *memory functions* introduced in GH are *ad hoc* mathematical expressions built with the aim of adapting the hydrodynamic equations and their solutions to the intermediate case of the viscoelastic regime of liquids, as in a typical "top-down" approach. This theoretical framework is normally referred to as the "memory functions formalism", or also the Zwanzig-Mori formalism [77-78, see also 7 and references therein].

The experimental evidences shortly described above may be summarized as follows:

1. propagation of light in liquids is affected by Doppler shift;

2. specific heat of monatomic liquids shows a temperature dependence typical of a substance in which a crystalline structure is present; it decreases from about $3\mathcal{R}$ at the melting point down to $2\mathcal{R}$ at the critical point;

3. when liquids are kept in controlled isothermal conditions, dynamic collective modes, both longitudinal and transversal, have been detected at very-high-frequencies, i.e. very-short-wavelengths;

4. these modes have been found to propagate in liquids at speeds close to that of ordinary sound in the corresponding solid, correlation lengths being higher than those normally attributed to liquids, which behave as a rigid network on such mesoscopic scale;

5. non-isothermal experiments performed in pure liquids and mixtures far from critical conditions have shown that the symmetry breaking induces long-range correlation fluctuations, increasing in turn the correlation lengths.

We will show in the following sections that the above experimental results may be interpreted within the DML. In this model, liquids are modelled as made of solid-like clusters in dynamic continuous re-arrangement, wandering in an amorphus ocean. Clusters, or "*liquid particles*", interact with the *lattice particles*, a population of elastic wave-packets (throughout this paper we will use the terms wave-packets, phonons, collective excitations, interchangeably), exchanging with them energy and momentum. Let us then proceed to describe this model.

## 3. The Dual Model of Liquids (DML)

This paragraph is made of four sections. In the first a description is provided of how the wave-packets work as thermal excitations and carriers of energy and momentum in the DML. The expression for the interaction term is introduced. In the second section an expression for the gradient of elastic pressure is derived based on "first principles" of wave mechanics. It is shown that it consists of a thermal pressure gradient, giving rise to an inertial term, the same introduced in the previous section, working when the characteristics of the medium change, continuously or abruptly, along the wave propagation direction. In the third section the *lattice particle*↔*liquid particle* interaction is described at the base of the DML, that is built without an external gradient applied to the liquid; in the last section the pivotal role of relaxation times is analyzed, their OoM is calculated and compared with the experimental data, as well the consequences of the model on the energy associated with the collective modes duly analysed.

### *3.1    Thermal excitations in liquids*

The large differences in transport phenomena between solids and gases are due to the different underlying mechanisms of molecular interactions. In gases interactions among molecules are very weak, resulting in random motions and slow diffusion. In solids the molecules are so strongly bonded to the lattice sites that gradient of concentration or moderate gradient of electrical potential cannot dislodge them. Solids however exhibit elevated thermal conductivities, especially isotropic pure crystals, such as diamond. This behaviour may be explained by the Debye model of solids, in which thermal (and elastic) energy is transported by waves (phonons). In a perfect solid lattice, in which the potential energy is quadratic in the distance from the equilibrium positions, elastic and thermal perturbations are carried by harmonic waves, able as such to transport only energy. Of course heat diffusion in solids, including perfect crystals, is not instantaneous and anharmonic terms arise because the potential energy is not exactly quadratic. This picture may be applied also to the liquid state, where the number of anharmonicities is much larger than in solids, as confirmed by the difference in the Grüneisen constant of various substances in the solid and liquid states. The fundamental role of anharmonicities consists in make it possible to exchange also momentum other than energy, providing thus an explanation for the diffusion and thermal diffusion in liquids.

The value of $\approx 3200\, m/s$ for the speed of sound in liquid water measured on mesoscopic scale up to distances of several molecular diameters is one of the driving argument of the DML, in which liquids are modelled as constituted by dynamic clusters, i.e. solid-like aggregates of molecules of the liquid in continuous re-arrangement, fluctuating and interacting with the liquid global system. Solid consists of only a crystalline, or pseudo-crystalline, phase. As the temperature increases, we could figure a non instantaneous melting, the solid phase giving progressively way to clusters swimming in an ocean of amorphous liquid. As far as a cluster exists, liquid molecules are bonded to the local lattice of the cluster to which they belong; the number and size of these domains of coherence decrease and the amount of amorphous liquid increases, to the point where it reaches the pure liquid (at the historically-called Frenkel line). As far as thermal (elastic) perturbations propagate within such a cluster, they behave as in solids. When perturbations cross the boundary between two such local lattices[2], inertial effects develop and the interactions are accompanied by propagation of elastic energy in forms of wave-packets because of the anharmonicity of the potential field arising at the cluster' border. Anharmonicities make the interaction time non negligible allowing for momentum transport which gives rise to the displacement of the icebergs, i.e. to their diffusion (Figure 1) [36-40,42-43,79]. The elastic waves travel across the solid-like local lattice at speeds typical of the corresponding solid phase; when the perturbations cross the boundaries of the local domains, their speed decreases and the other liquid parameters undergo the same fate. This arrangement by local lattices of liquid molecules on mesoscopic scale justifies the experimental value for the speed of sound in water (and in other liquids) close to that of the ice, as well the PSD $du^{wp}/d\nu$ of sound velocity observed in liquids with respect to the frequency [19-21,24-26].

What one usually means for a "liquid" is seen in the DML as a mixture of solid-like clusters of molecules in continuous re-arrangement and an amorphous, disordered liquid phase. As consequence, any liquid parameter experimentally measured at temperatures and pressures where a usual liquid phase exists, is actually a pondered average of a solid/liquid value.

Let's now translate the ideas illustrated above into a model of the liquid state using the laws of physics. Figure 3 represents the "*liquid particle* ↔ *wave-packet*" interaction, and shows how these interactions work. Because the interaction with the *liquid particle* involves wave-packets and not harmonic waves, it takes a finite time $\langle \tau_p \rangle$, during which the cluster is displaced by $\langle \Lambda_p \rangle$ (here and in the rest of the paper, the two brackets $\langle \ \rangle$ indicate the average over a statistical ensemble of the quantity inside them). The interaction allows exchanging of both energy and momentum between the phonon and the *liquid particle*. In an event of type a) an energetic wave-packet collide with a *liquid particle*, transferring to it the energy $\Delta \varepsilon^{wp}$ and momentum $\Delta p^{wp}$, given by:

4. $\quad \Delta \varepsilon^{wp} = h \langle v_1 \rangle - h \langle v_2 \rangle = \Delta E_p^k + \Delta \Psi_p = f^{th} \cdot \langle \Lambda_p \rangle$

5. $\quad \Delta p^{wp} = f^{th} \cdot \langle \tau_p \rangle.$

In Eq.(4) $\langle v_1 \rangle$ and $\langle v_2 \rangle$ represent the wave-packet frequency before and after the collision, respectively. $\Delta E_p^k$ is the kinetic energy acquired by the *liquid particle*, while $\Delta \Psi_p$ is that part of the energy lost by the wave-packet in the inelastic interaction and converted into potential energy of internal collective DoF of the solid-like cluster. Finally $f^{th}$ is the interaction term between the wave-packet and the *liquid particle*. As consequence of the collision, the wave-packet loses energy and momentum, while the cluster acquires the energy and momentum lost by the former. The effect is of having converted the energy carried by the wave-packet into kinetic and potential energy of the cluster. $W = f^{th} \cdot \dfrac{\langle \Lambda_p \rangle}{\langle \tau_p \rangle}$ is the power developed during the interaction. The kinetic energy acquired by the *liquid particle* will be dissipated as friction against the liquid. The *liquid particle* will begin to relax the potential energy $\Delta \Psi_p$ once the interaction is completed, i.e. after $\langle \tau_p \rangle$, and will dissipate it during the time lapse $\langle \tau_R \rangle$, travelling over the distance $\langle \Lambda_R \rangle$, at the end of which the residual energy stored into internal DoF returns to the liquid pool. The total displacement of the *particle* and the total duration of the process are:

6. $\quad \langle \Lambda \rangle = \langle \Lambda_P \rangle + \langle \Lambda_R \rangle$

7. $\quad \langle \tau \rangle = \langle \tau_P \rangle + \langle \tau_R \rangle$

---

[2] In case of solutions, this same behaviour shall be observed among a cluster of solvent and a solute particle; solute particles of course will have a more complicate structure, but however comparable to that of a cluster of pure solvent from the point of view of the heat current propagation.

An interesting observation is that the interaction looks like a tunnel effect, inasmuch the energy subtracted from the phonon' pool returns into it a time interval $\langle \tau \rangle$ later and a step $\langle \Lambda \rangle$ forward. The event of type b) is the time-reversal of a): an energetic cluster interact with a wave-packet transferring to it energy and momentum, In this case the outgoing wave-packet has larger energy and momentum and the net effect is of having increased the liquid thermal energy carried by phonons at expenses of the kinetic and internal energy of the *liquid particle* (interestingly, $\langle \Lambda \rangle$ and $\langle \tau \rangle$ in Eqs.(6) and (7) have the same meaning as $\delta$ and $\tau_F$ defined by Frenkel [5], (see Figure 2)). The tunnel effect in this case consists in shifting a *liquid particle* by $\langle \Lambda \rangle$ from one place to another. In equilibrium conditions, events a) will alternate over time with events b), to keep unchanged in the average the balance of the two energy pools. The mesoscopic equilibrium induces the macroscopic equilibrium: events like those of Figure 3 will be equally probable along any direction, with a null average over time and space. In case the symmetry is disrupted by an external force field, as for instance by a temperature or a concentration gradient, one type of event will prevail over the other along the direction of the gradient. The *lattice particle* ↔ *liquid particle* interaction has the additional feature of giving a propagative character to the advancement of the thermo-elastic wave rather than diffusive, with propagation velocity:

8. $\quad \langle v_s \rangle = \dfrac{\langle \Lambda \rangle}{\langle \tau \rangle}$

We want observe that, while the kinetic energy is a continuum, the internal vibratory collective DoF are quantized. This means that to excite them, the energy transfer must follow the rules of quantum theory, i.e. only finite quanta of energy can be exchanged between the internal DoF and the external pool; the only possible exception is that of anharmonic oscillators that can generate a wide spectrum of frequencies in the interaction process, as hypothesized in DML. Besides, as we will see in the *Discussion*, an exchange of energy between internal (vibratory) and external (translatory) DoF is made possible under precise circumstances.

The interactions described in Figure 3 have the following characteristics:

a) Both energy and momentum transport are allowed, the collisions being inelastic.

b) Everyone of the two events a) and b) is commuted into the other by time reversal. At equilibrium, the average energy and momentum exchanged in events of type a) are equal to those exchanged in events of type b); in other words, at equilibrium the two events are statistically equivalent and have the same *a priori* probability to occur. If a symmetry breaking is imposed to the system, as a temperature or a concentration gradient, one of the two events will have a larger probability to occur with respect to the other.

c) Part of the energy transferred by the wave-packet to the *liquid particle* increases the energy of internal DoF. This is shown in Figure 3a), where on the left there is a non perturbed *liquid particle* just before the collision with a wave-packet takes place. The impact transfers part of the energy of the wave-packet to the cluster; this energy is partly converted into kinetic energy of the cluster, and partly into potential energy, exciting the internal vibratory DoF (middle of the figure). The collision lasts $\langle \tau_p \rangle$ and the *liquid particle* is displaced by $\langle \Lambda_p \rangle$. Once the collision is completed and the phonon has emerged with reduced energy and momentum, the *liquid particle* de-excites and releases the excess energy in the time interval $\langle \tau_R \rangle$, moving by $\langle \Lambda_R \rangle$ (right side of the figure). In Figure 3b) the time-reversed event of a) is represented.

d) The elementary interaction between wave-packet and liquid particle has two other very interesting implications; i) it gives a propagation character to the advancement of the thermo-elastic wave rather than a diffusive one, representing the mechanism by which energy is displaced from one place to another, alike in a tunnel effect; the ratio between $\langle \Lambda \rangle$ and $\langle \tau \rangle$ of Eqs. (6) and (7) is the wave propagation velocity $\langle v_s \rangle$; ii) it provides an elementary model at molecular level for the viscosity in liquids, representing the mechanism by which the medium particle exchange momentum with the lattice.

One of the possible effects which one would get by accounting for the energy stored in non-propagating modes will not be dealt with here. It has indeed an impact on the heat propagation, a topic that will be accurately dealt with in a separate paper [80].

*3.2  The Radiant Vector and the elastic radiation pressure*

Having stated that thermal excitations in liquids consist of elastic wave-packets, we now calculate the pressure generated by the flux of elastic energy travelling in a condensed medium. This elastic radiation pressure generates the interaction term $f^{th}$ responsible of the *wave-packet* ↔ *liquid particle* interaction introduced in Eqs. (4) and (5).

Energy density *E* associated with longitudinal mechanical waves consists of a kinetic and a potential part:

9.     $E = \langle E_{kin} \rangle + \langle E_{pot} \rangle = \rho \langle \dot{\xi}^2 \rangle$

$\rho$ being fluid density and $\dot{\xi}$ oscillation velocity of the particles. The flux of elastic energy $\vec{J}_{el} = (E \cdot u_g)\vec{r}$ where $u_g$ is group velocity and $\vec{r}$ the unit vector along the direction of propagation. It is also true:

10. $$\vec{J}_{el} = (\Delta\Pi \cdot \dot{\xi})\vec{r} = \vec{R}$$

$\Delta\Pi$ being the pressure variation. The quantity $\vec{R}$, which we call *radiant vector* [81-82], represents the instantaneous acoustic power in the beam; it is never negative because $\Delta\Pi$ and $\dot{\xi}$ change sign simultaneously every half period. Dividing Eq. (10) by the phase velocity $u_\varphi$, one gets the elastic radiation pressure that manifests in a medium along the elastic wave propagation direction, according to the classic Rayleigh approach. If the characteristics of the medium change, a gradient of elastic pressure rises, $\nabla\Pi_{el} = \nabla\left(\dfrac{E \cdot u_g}{u_\varphi}\right) = \nabla\left(\dfrac{J_{el}}{u_\varphi}\right)$.

The reader should note that the concept of *radiant vector* holds for longitudinal waves of any frequency and amplitude. As shown in the previous section, elastic wave-packets are also responsible for propagation of heat waves inside condensed media. What was said for the flux of elastic energy can also be said for the flux of thermal energy, as also pointed out by Joyce [82]. We replace the gradient of elastic pressure with the gradient of thermal radiation pressure, and the flux of elastic energy with that of thermal energy, yielding:

11. $$\nabla\Pi_{th} = \nabla\left(\dfrac{J_q}{u_\varphi}\right) = -\varphi^{th}$$

Eq.(11) contains two supplementary information; first, because thermal energy is carried by elastic wave-packets, the velocity of energy propagations remains the same. Second, the gradient of pressure gives rise to an inertial term, $\varphi^{th}$, responsible for momentum transport associated with the wave-packets propagation. Eq.(11) is valid when the characteristics of the medium change continuously along the wave-packets propagation direction. The same argument may be used when wave-packets impinge on an "obstacle", as for instance a *liquid particle*: the *radiant vector* changes and an acoustic radiation pressure is produced on the boundary. Observing that the inelastic character of the collision allows exchange of momentum, this leads to the appearance of a force, $f^{th}$, acting on molecular clusters. Consequently, we rewrite Eq.(11) in terms of discrete quantities, yielding the following expression for the thermal radiation pressure:

12. $$\Delta \Pi_{th} = \left[\left(\frac{J_q}{u_\varphi}\right)_1 - \left(\frac{J_q}{u_\varphi}\right)_2\right] = \frac{f^{th}}{\sigma_p}.$$

Wave-packets exchange energy $\Delta\varepsilon = f^{th} \cdot \langle \Lambda_p \rangle$ and momentum $\Delta p = f^{th} \cdot \langle \tau_p \rangle$ with *liquid particles*, so that change of local order produces at the mesoscopic scale on the wave-packets the same effect of a boundary, resulting in a change of their radiant vectors.

Expressions analogous to Eq.(12) have been previously derived by many Authors in various different approaches. It derives from the Boltzmann-Ehrenfest' Adiabatic Theorem [83], later generalized by Smith [84] and by Gaeta et al. [85]; in a different approach analogous result was obtained calculating the radiation pressure produced by acoustic waves on an idealized liquid-liquid interface [41].

The third member of Eq.(12) provides the expression for $f^{th}$ introduced in Eqs.(4) and (5),

13. $$f^{th} = \sigma_p \left[\left(\frac{J_q}{u_\varphi}\right)_1 - \left(\frac{J_q}{u_\varphi}\right)_2\right]$$

$\sigma_p$ representing the cross-section of the "obstacle", the solid-like cluster, on the surface of which $\Delta\Pi_{th}$ develops. A final consideration concerns the algebraic sign of $f^{th}$, or that of $\Delta\Pi_{th}$ or $\Delta\Pi_{el}$: it will be positive or negative depending on the sign of the quantity in square brackets, i.e. on whether the energy associated with the propagation of elastic waves in the medium "1" is larger or smaller than that in the medium "2".

### 3.3 *Lattice particle ↔ liquid particle interactions and thermal energy associated with collective modes in the DML*

The existence in liquids of thermal excitations, consisting of high-frequency wave-packets, capable of interacting with the molecules of the liquid, has many interesting consequences. Let's start by considering the propagation of thermal (elastic) energy in liquids due to wave-packets diffusion. The total internal energy $q_T$ per unit of volume of a liquid at temperature T is:

14. $$q_T = \int_0^T \rho C_V d\theta = f[\Theta/T]$$

where $\rho$ is the medium density, $C_V$ the specific heat per unit mass at constant volume, $\Theta$ the Debye temperature of the liquid at temperature T. The fraction $q_T^{wp}$ of this energy is transported by the wave-packets:

15. $\quad q_T^{wp} = m q_T = \mathcal{N}^{wp} \langle \varepsilon^{wp} \rangle = m^* \rho C_V T$

The parameter $m$ is introduced to account for the involvement of the DoF of the lattice; $m$ is the ratio between the number of collective DoF surviving at temperature T, and the total number of available collective DoF, then it holds $0 \leq m \leq 1$. $\mathcal{N}^{wp}$ is the number of wave-packets per unit of volume[3], and $\langle \varepsilon^{wp} \rangle$ their average energy; the quantity $m^* = m \dfrac{\int_0^T \rho C_V d\theta}{\rho C_V T}$ depends on the nature of the liquid, and is introduced to simplify the expression.

In the absence of external perturbation to the system the wave-packets will be randomly and isotropically oriented. With reference to Figure 1 and Figure 3, during their free-flight $\langle \Lambda_{wp} \rangle$ between two successive interactions with the *liquid particle*, the wave-packets may be considered as local microscopic heat currents. Each *lattice particle* propagates along an average distance $\langle \Lambda_{wp} \rangle$, during a time interval $\langle \tau_{wp} \rangle$, which is its average life-time. Accordingly, the ratio $\langle \Lambda_{wp} \rangle / \langle \tau_{wp} \rangle$ defines the wave-packet average propagation velocity between two successive interactions,

16. $\quad u^{wp} = \dfrac{\langle \Lambda_{wp} \rangle}{\langle \tau_{wp} \rangle} = \lambda^{wp} \cdot \nu^{wp}$.

where $\nu^{wp}$ and $\lambda^{wp}$ are the central frequency and wave-length of the wave-packet[4]. Considering a surface S arbitrarily oriented within the liquid, at equilibrium an equal number of wave-packets will flow through it in opposite directions. Let's assume now that locally every heat current, although isotropically distributed in the liquid, is driven by a *virtual temperature gradient* $\left\langle \frac{\delta T}{\delta z} \right\rangle$ applied over $\left\langle \Lambda_{wp} \right\rangle$. From Eq.(15), the variation of the energy density within the liquid along the "+z" direction is:

17. $\quad \delta_{+z} q_T^{wp} = \frac{1}{6} \delta(m q_T) = \frac{1}{6} \delta\left[ \mathcal{N}^{wp} \left\langle \varepsilon^{wp} \right\rangle \right] = \frac{1}{6} \mathcal{N}^{wp} \left\langle \Delta \varepsilon^{wp} \right\rangle = \frac{1}{6} \frac{\partial q_T^{wp}}{\partial T} \left\langle \frac{\delta T}{\delta z} \right\rangle \left\langle \Lambda_{wp} \right\rangle$

---

[3] $\mathcal{N}^{wp}$ is the average density of the wave-packets statistical distribution. As such $\mathcal{N}^{wp}$ is represented by a Bose-Einstein distribution function.

[4] Although an elastic perturbation of the liquid lattice should be represented by a wave-packet, whose localization depends on the microscopic structure to which it is associated, here it will be represented, for the sake of mathematical simplification, before and after an interaction, as a monochromatic wave, longitudinally polarized, and hence identified by its average wave-length $\lambda^{wp}$ and frequency $\nu^{wp}$, whose product is equal to $u^{wp}$.

In the third to fourth member step of Eq.(17) it is supposed that the wave-packets distribution function $\mathcal{N}^{wp}$ is not significantly altered along the heat current, and that only their average energy changes. The local virtual temperature gradient $\left\langle \frac{\delta T}{\delta z} \right\rangle$ in last member represents the thermodynamic *"generalized"* force [with the usual meaning of irreversible thermodynamics] driving the diffusion of thermal excitations along *z*. The factor $1/6$ accounts for the isotropy of the phonon current through a liquid at equilibrium. By applying this driving force over the distance $\left\langle \Lambda_{wp} \right\rangle$, a wave-packet diffusion, i.e. the heat current $j^{wp}$, is generated along the "+z" direction:

18. $\quad j_{+z}^{wp} = -D^{wp} \left( \frac{\delta q_T^{wp}}{\delta z} \right)_{+z} = -\frac{1}{6} u^{wp} \left\langle \Lambda_{wp} \right\rangle \frac{\partial q_T^{wp}}{\partial T} \left\langle \frac{\delta T}{\delta z} \right\rangle$

where $D^{wp} = u^{wp} \left\langle \Lambda_{wp} \right\rangle$ is the diffusion coefficient of the wave packets. Therefore, the propagation of an elementary heat current lasts the time interval $\left\langle \tau_{wp} \right\rangle$ and is driven by a virtual temperature gradient $\left\langle \frac{\delta T}{\delta z} \right\rangle$ over the distance $\left\langle \Lambda_{wp} \right\rangle$.

Eq. (18) for the heat current may also be obtained by calculating the energy flowing in a cylinder per unit cross section and per unit time:

19. $\quad j_{+z}^{wp} = u^{wp} \delta_{+z} q_T^{wp} = \frac{1}{6} u^{wp} \mathcal{N}^{wp} \left\langle \Delta \varepsilon^{wp} \right\rangle$.

$q_T^{wp}$ in Eq.(15) allows to calculate the phonon specific heat contribution $C_V^{wp}$:

20. $\quad \begin{cases} \rho C_V^{wp} = \dfrac{\partial q_T^{wp}}{\partial T} = \dfrac{\partial}{\partial T}(m q_T) = q_T \dfrac{dm}{dT} + m \dfrac{\partial q_T}{\partial T} = q_T \dfrac{dm}{dT} + m \rho C_V = \\ = m \rho C_V \left[ \dfrac{q_T}{m \rho C_V} \dfrac{dm}{dT} + 1 \right] = m \rho C_V \left[ \dfrac{m^*}{m^2} \dfrac{dm}{dT} T + 1 \right] \end{cases}$

Let's evaluate now the quantity in square brackets at the last member. Because

21. $\rho C_V^{wp} = \dfrac{\partial q_T^{wp}}{\partial T} \geq 0$,

and $0 \leq m \leq 1$, this implies that $\left[\dfrac{m^*}{m^2}\dfrac{dm}{dT}T + 1\right] \geq 0$.

The number of collective DoF defined by $m$ decreases with temperature, so that $\dfrac{dm}{dT} < 0$. Consequently, $\dfrac{m^*}{m^2}\dfrac{dm}{dT}T \geq -1$, and this defines the lower limit for $\dfrac{m^*}{m^2}\dfrac{dm}{dT}T$. As for the upper limit, because of the fact that $\dfrac{dm}{dT} < 0$ and $\left[\dfrac{m^*}{m^2}\dfrac{dm}{dT}T + 1\right] \geq 0$, we get for the maximum value $\dfrac{m^*}{m^2}\dfrac{dm}{dT}T \leq 0$. Definitively we have:

22. $0 \leq \left[\dfrac{m^*}{m^2}\dfrac{dm}{dT}T + 1\right] \leq 1$

and finally

23. $C_V^{wp} \leq C_V$

as one would have expected.

As for the temperature dependence of *m*, we will discuss in Section 5 how this parameter evolves in liquids from the glass transition up to the Frenkel line, and how it is modelled in the DML. The basis for the reasoning is the experimental evidence of the presence in liquids of transversal modes [21,23-29,31-35,59,65-66] actives for the propagation of elastic energy by means of shear waves working as in the solid phase. These modes however persist in the liquid phase as long as solid-like structures survive. Experiments show that the two transversal modes disappear when the system approaches the critical point, where only the longitudinal collective modes survive accounting for the compression and rarefaction waves responsible for hydrodynamic modes propagation.

Besides being an energy density, $q_T^{wp}$ represents also the pressure $\Pi^{wp}$ exerted by elastic wave-packets in the medium. Compiling Eq. (12) with Eqs.(17), (18) and (20), along "+z" we get:

24. $\Pi_{+z}^{wp} \equiv \delta_{+z} q_T^{wp} = \dfrac{1}{6}\delta q_T^{wp} = \dfrac{1}{6}\delta\left(\dfrac{j^{wp}}{u^{wp}}\right) = \delta_{+z}\left(\dfrac{j^{wp}}{u^{wp}}\right) = \dfrac{\partial q_T^{wp}}{\partial T}\left\langle\dfrac{\delta T}{\delta z}\right\rangle_{+z}\langle\Lambda_{wp}\rangle = \rho C_V^{wp}\langle\Lambda_{wp}\rangle\left\langle\dfrac{\delta T}{\delta z}\right\rangle_{+z}$

Let us now return to Eq.(18), which represents the flux of thermal energy carried by a wave-packet propagating over the distance $\langle \Lambda_{wp} \rangle$ and driven by the virtual gradient $\langle \frac{\delta T}{\delta z} \rangle$. $j_{+z}^{wp}$ may be re-written introducing the (mesoscopic) thermal conductivity of the medium $K^{wp}$:

25. $\quad j_{+z}^{wp} = -K_{+z}^{wp} \langle \frac{\delta T}{\delta z} \rangle_{+z} = -\frac{1}{6} \{ u^{wp} \langle \Lambda_{wp} \rangle \rho C_V^{wp} \} \langle \frac{\delta T}{\delta z} \rangle = -\frac{1}{6} D^{wp} \rho C_V^{wp} \langle \frac{\delta T}{\delta z} \rangle$

from which we get:

26. $\quad K_{+z}^{wp} = \frac{1}{6} D^{wp} \rho C_V^{wp} = \frac{1}{6} u^{wp} \langle \Lambda_{wp} \rangle \frac{\partial}{\partial T}[\mathcal{N}^{wp} \langle \varepsilon^{wp} \rangle] = \frac{1}{6} u^{wp} \langle \Lambda_{wp} \rangle m \rho C_V \left[ \frac{m^*}{m^2} \frac{dm}{dT} T + 1 \right]$

What stated so far for the direction "$+z$" holds analogously for the opposite direction "$-z$", getting the same expression as Eq.(26) but with the opposite sign; by summing them algebraically, we eventually get:

27. $\quad K^{wp} = \frac{1}{3} u^{wp} \langle \Lambda_{wp} \rangle \frac{\partial}{\partial T}[\mathcal{N}^{wp} \langle \varepsilon^{wp} \rangle] = \frac{1}{3} u^{wp} \langle \Lambda_{wp} \rangle m \rho C_V \left[ \frac{m^*}{m^2} \frac{dm}{dT} T + 1 \right] = \frac{1}{3} u^{wp} \langle \Lambda_{wp} \rangle \rho C_V^{wp}$

Interestingly, Eq. (27) for the (phononic) thermal conductivity has exactly the same form as that derived by Boltzmann for the thermal conductivity due to the particle collision and drift in a fluid medium [86, Eq.(86)]. In particular, the quantity $G$ introduced by Boltzmann is here replaced by $\rho C_V^{wp}$, that in turn depends upon the number of DoF pertaining to the *particles* involved in the process [86, Eq.(88)].

So far the physical implications on the dynamics of pure liquids at equilibrium (the local temperature gradient $\langle \frac{\delta T}{\delta z} \rangle$ is indeed a *virtual* quantity accounting for the wave-packets current) deriving from the DML. The main conclusions arrived at are the following. In the DML $f^{th}$ develops when the characteristics of the medium change; this happens in pure liquids at equilibrium at the boundary of the local ordered domains (i.e. the *liquid particles*) within which liquids may be considered as solids as long as these domains of coherence last. With reference to Figure 3a, when the wave-packet hits a *liquid particle*, the force given by Eq.(13) develops, it acts for $\langle \tau_p \rangle$ displacing the *particle* by $\langle \Lambda_p \rangle$. Energy $\langle \Delta \varepsilon^{wp} \rangle$ and momentum $\langle \Delta p^{wp} \rangle$, given by Eqs.(4) and (5) respectively, are transferred to the *particle* during $\langle \tau_p \rangle$, increasing its kinetic and potential energy and exciting internal vibratory energy levels. These DoF oscillate similarly to those pertaining to solid state, giving origin to (quasi) elastic waves with characteristic

length- and time-scales $\langle\Lambda_0\rangle$ and $\langle\tau_0\rangle$ respectively. $\langle\tau_R\rangle$ later and $\langle\Lambda_R\rangle$ forward they return to the wave-packet pool alike in a tunnel effect; the process is repeated every time interval $\langle\tau\rangle$ and every step $\langle\Lambda\rangle$. Because the events b) are the time-reversal of a), their statistical equilibrium ensures that no macroscopic effects are developed.

When a symmetry breaking is imposed to the system, as for instance a temperature (or concentration) gradient, there will be no more statistical equilibrium of the events of Figure 3, and an excess of events a) (or b)) will result along the direction of the externally applied (generalized) force. The local heat currents will generate forces and observable effects of these microscopic events will appear in the form of energy (mass) flux. However, the evaluation of the dynamical effects from Eq.(13) in a traditional liquid is not a simple exercise because a "*pure liquid phase*" exists only at temperatures close to the so-called Frenkel line. In Eq.(13), or Eq.(12), it is therefore difficult to identify which values to use if one wishes to evaluate it in a pure liquid or in a solution. Aware of making a mistake, an evaluation of OoM can be made by assuming, in a pure liquid, for the clusters the values of the solid and for the amorphous phase those of the liquid.

## *3.4 Relaxation times Order-of-Magnitude evaluation in the DML and comparison with experiments on viscoelastic transition regions*

Let us then now analyze the relaxation time $\langle\tau\rangle$ that one should consider in the frame of DML to account for the delay in the energy (or mass) diffusion. To do this we shall refer to Figure 4 that represents a close-up picture of the "*liquid particle ↔ lattice particle*" interaction described in Figure 3a. Being wave-packets and not harmonic waves, the inelastic interaction with the liquid particle takes a finite time $\langle\tau_p\rangle$, during which the particle is displaced by $\langle\Lambda_p\rangle$, both energy and momentum being exchanged between the phonon and the particle. In an event of type a) an energetic wave-packet collide with a solid-like *liquid particle*, transferring to it energy $\Delta\varepsilon^{wp}$ and momentum $\Delta p^{wp}$, given by Eqs. (4) and (5) respectively. $\Delta E_p^k$ in Eq. (4) is the kinetic energy acquired by the *liquid particle* as consequence of the interaction, and given by:

$$28.\quad \Delta E_p^k = \frac{1}{2}m_p\left(\frac{\langle\Lambda_p\rangle}{\langle\tau_p\rangle}\right)^2 = \frac{1}{2}m_p\upsilon_p^2$$

where $m_p$ is the mass of the displaced *liquid particle*.

The particle will begin to relax the potential energy $\Delta\Psi_p$ once the interaction is completed, i.e. after $\langle\tau_p\rangle$, dissipating it during a time lapse $\langle\tau_R\rangle$, and travelling over a distance $\langle\Lambda_R\rangle$, at the end of which the energy stored into internal DoF returns to the liquid pool. The total duration of the process and the total displacement of the particle are given by Eqs. (6) and (7), respectively[5]. Because a wave-packet is not an infinite wave train but a localized perturbation with extension $\langle\Lambda_{wp}\rangle$, we may write it as a multiple of $\lambda^{wp}$ [Eq.(16)]:

29. $\langle\Lambda_{wp}\rangle = n\lambda^{wp}$, $n > 1$.

In Figure 4 $\langle d_p\rangle$ indicates the cross extension of the *liquid particle*; the interaction starts when the phonon collide with the *particle*, and ends when the phonon leaves the particle, after having crossed it at the speed

30. $u^0 = \dfrac{\langle\Lambda_0\rangle}{\langle\tau_0\rangle}$.

During the interaction, lasting $\langle\tau_p\rangle$, the wave-packet transfers to the *liquid particle* a non-negligible part of its total energy, $\phi = \dfrac{h\langle v_1\rangle - h\langle v_2\rangle}{h\langle v_1\rangle}$. A fraction $\alpha$ of this becomes kinetic energy of the iceberg, while the remaining fraction $1-\alpha$ becomes potential energy $\Delta\Psi_p$ of internal DoF (quantized vibratory DoF):

31. $\Delta E_p^k = \alpha \cdot h\langle v^{wp}\rangle = \dfrac{1}{2}m_p\langle v_p^2\rangle \Rightarrow \langle v_p\rangle = \sqrt{\dfrac{2\alpha h\langle v^{wp}\rangle}{m_p}}$

From Figure 4, the total path of the phonon $\langle\Delta^{ph}\rangle$ is:

32. $\langle\Delta^{ph}\rangle = \langle\Lambda_{wp}\rangle + \langle d_p\rangle + \langle\Lambda_p\rangle + \langle\Lambda'_{wp}\rangle$

where the quantity $\langle\Lambda'_{wp}\rangle$ is the extension of the wave-packet after the interaction, with (central) frequency $\langle v_2\rangle$. Of course the total duration of the interaction experienced by the particle is the same as that experienced by the phonon, but because they travel at different speeds, the covered paths will be different. For the only sake of distinguishing among them, we indicate with $\langle \tau_p \rangle$ the duration of the interaction for the particle, and with $\langle \tau^{ph} \rangle$ that for the phonon, with $\langle \tau_p \rangle \equiv \langle \tau^{ph} \rangle$:

33. $\langle \tau_p \rangle = \dfrac{\langle \Lambda_p \rangle}{\langle \upsilon_p \rangle}$

34. $\langle \tau^{ph} \rangle = \dfrac{\langle \Lambda_{wp} \rangle}{u^{wp}} + \dfrac{d_p}{u^0} + \dfrac{\langle \Lambda_p \rangle}{u^0} + \dfrac{\langle \Lambda'_{wp} \rangle}{u^{wp}}$

---

[5] Actually the time interval between two successive interactions, $\tau_F$, may be larger than $\langle\tau\rangle$ because another time interval could elapse after the particle is stopped or has dissipated the internal energy acquired during the interaction.

In Eq.(34), the first term at the second member is the time taken by the wave-packet to enter the *liquid particle*, the second is the time taken by the phonon to cross the liquid particle, the third is the time taken to cover the entire path of interaction, and finally the fourth term is the time taken by the wave-packet to leave the *liquid particle*. Two approximations are in order for Eq.(34). First, because the emerging wave-packet has a reduced energy with respect to the colliding one, the wavelength of the former will certainly be larger than the latter; we use however for it the same extension as the colliding one, $\langle \Lambda_{wp} \rangle$. The second approximation concerns the speed at which the colliding and emerging wave-packets enter and leave the *liquid particle*. We use the same speed for both, and equal to the speed of elastic waves inside the liquid particle, $u^0$, for which, as described in Section 2, the value is known from experiments. Ultimately, Eq.(34) becomes:

35. $\langle \tau^{ph} \rangle \cong \dfrac{2\langle \Lambda_{wp} \rangle + d_p + \langle \Lambda_p \rangle}{u^0}$

From Eqs.(32) through (35) we have:

36. $\dfrac{\langle \Delta^{ph} \rangle}{\langle \Lambda_p \rangle} \cong \dfrac{2\langle \Lambda_{wp} \rangle + \langle d_p \rangle + \langle \Lambda_p \rangle}{\langle \Lambda_p \rangle} = \dfrac{2\langle \Lambda_{wp} \rangle + \langle d_p \rangle}{\langle \Lambda_p \rangle} + 1 = \dfrac{u^0}{\langle \upsilon_p \rangle}$

from which, considering the last two members,

37. $\dfrac{2\langle \Lambda_{wp} \rangle + \langle d_p \rangle}{\langle \Lambda_p \rangle} = \dfrac{u^0}{\langle \upsilon_p \rangle} - 1$

and finally

38. $\langle \Lambda_p \rangle = \dfrac{\langle \upsilon_p \rangle \left( 2\langle \Lambda_{wp} \rangle + \langle d_p \rangle \right)}{u^0 - \langle \upsilon_p \rangle}$

Compiling Eqs.(31), (34) and (38) it is now possible to get $\langle \tau_p \rangle$ (or $\langle \tau^{ph} \rangle$).

It is easy to evaluate also the time $\langle\tau_\eta\rangle$ a liquid particle takes to stop after the interaction, by supposing that the kinetic energy acquired is dissipated against friction in the liquid, $\Delta E_p^k = E_p^\eta$. Although $\langle\tau_\eta\rangle$ has nothing to do with $\langle\tau_R\rangle$, it can tell us how long after the interaction the particle stops and how much space it travels before a successive interaction, provided that no other collisions have occurred during its flight. To answer such doubt we may get help evaluating the OoM of the phonon density $\mathcal{N}^{wp}$ that can be deduced from Eq.(14); it provides the variation of the energy density $\delta q_T^{wp} = \delta(mq_T)$ within the liquid along a given direction. This energy variation is equal to the energy shift associated to a Brillouin doublet [Eq.(1)] multiplied by the phonon density, $\mathcal{N}^{wp}\langle\Delta\varepsilon^{wp}\rangle$. Definitively, in water, with a value of 1 calorie for the specific heat and with the same values for the wave-packet frequency as those given above, one has:

39. $$\mathcal{N}^{wp}\langle\Delta\varepsilon^{wp}\rangle = \mathcal{N}^{wp}h\Delta\langle\nu^{wp}\rangle = \delta(mq_T) = m\rho C_V \Delta T \Rightarrow \mathcal{N}^{wp} = \frac{m\rho C_V \Delta T}{\langle\Delta\varepsilon^{wp}\rangle} \approx m \cdot 4 \cdot 10^{22}\, cm^{-3}$$

The OoM of the density of phonons, dealt with as *lattice particles*, in Eq.(39) is then of the same OoM as the molecular density (for water the molecular density holds $\approx 4 \cdot 10^{22}\, cm^{-3}$ at ambient temperature), or even lesser, because $0 \leq m \leq 1$. This result supports the hypothesis of neglecting multiple interactions[6] of wave-packets with particles. The numerical result of Eq.(39) is obtained assuming that the density of phonons does not change by changing the temperature, the only variation being that of their energy. Therefore the above considerations tell us that $\langle\tau_\eta\rangle$ represents a lower limit for the remaining time interval after $\langle\tau_p\rangle$ before another interaction with a phonon occurs, or before the liquid particle dissipates all of its internal energy. A dedicated analysis of the variation of $\mathcal{N}^{wp}$ with temperature or frequency in the object of a separate paper.

Using the Stokes-Einstein expression, the energy balance equation holds:

40. $$\frac{1}{2}m_p\langle\upsilon_p^2\rangle = 6\pi\eta_l\langle r_p\rangle\langle\upsilon_p\rangle\langle\Lambda_R\rangle \Rightarrow$$

41. $$\langle\upsilon_p\rangle = \frac{dx_p}{dt} = \frac{12\pi\eta_l\langle r_p\rangle}{m_p}x_p$$

where $\langle r_p\rangle = \langle d_p\rangle/2$, from which we get:

---

[6] The occurrence of multiple interactions can be excluded, as discussed below (see Eq.(39)).

42. $\dfrac{dx_p}{x_p} = \dfrac{dt}{\langle \tau_\eta \rangle}$ with $\langle \tau_\eta \rangle = \dfrac{m_p}{12\pi \eta_l \langle r_p \rangle}$

The distance travelled by the *liquid particle* is obtained by integrating Eq. (42), getting:

43. $x_p(t) = \langle \upsilon_p \rangle \langle \tau_\eta \rangle \left(1 - e^{-t/\langle \tau_\eta \rangle}\right)$.

To provide a numerical estimation of the OoM of the several variables introduced above, we suppose to be in water at *T=300K* as example. Considering several ranges for all the parameters has allowed us to evaluate the minimum and maximum values of the variables characterizing the dynamics of wave-packets and *liquid particles*. In doing this we have also taken into account the fact that, for a wave to "see" an obstacle and interact with it, its extension (or wavelength) must be less than, or at most comparable with, the size of the obstacle: $\langle \Lambda_{wp} \rangle \leq \langle d_p \rangle$. Results are provided in Table 1, from which several conclusions may be deduced. First, although $\langle \tau_\eta \rangle$ represents at the best only an OoM for $\langle \tau_R \rangle$, it is straightforward that the time interval dominating the dynamics is $\langle \tau_p \rangle$, by far much larger than $\langle \tau_\eta \rangle$. $\langle \tau_R \rangle$ is the average time taken by the internal (quantized) DoF of the iceberg to de-excite after a collision of the iceberg with a wave-packet has occurred. At present we have no tools to evaluate $\langle \tau_R \rangle$, being it however out of the physical context around which the DML is developed. A detailed quantum mechanical modelling of the internal DoF is the best suitable candidate able to evaluate $\langle \tau_R \rangle$. Anyway, we may certainly conclude from Table 1 that the OoM of total relaxation time is $\langle \tau \rangle > \langle \tau_p \rangle \approx 4 \div 13 \cdot 10^{-12} s$, i.e. of the order of several *ps,* as theoretically foreseen and experimentally deduced from scattering spectra.

Very interestingly, multiplying these values by the corresponding phonon frequencies used in the calculations, $\nu^{wp} \approx 0.9 \div 3 \cdot 10^{12} Hz$, we get:

44. $\nu^{wp} \cdot \tau \approx 3.6 \div 40$.

This result is in line with what foreseen from the PLT and deduced from experimental basis, thus confirming that in the considered ranges of temperature and pressure, the liquid is in a full viscoelastic regime, with a prominence towards the single particle, elastic regime at higher frequencies. This in turn is an indirect evidence of the presence of pseudo-crystalline, solid-like structures that manifest through the enhancement of the sound velocity, as well as of anharmonic interactions arising at the border of such structures.

## 4. Numerical Simulations

The liquid models, whatever their basis, are by nature non-Galilean systems, as they cannot in any way be directly tested or replicated in the laboratory, even in a simplified way. Such models, and those on liquids are no exception, today are evaluated on the basis of the Popperian criterion of non-falsifiability: when it is not possible to prove the truthfulness of a hypothesis, it can be accepted until proven falsifiable. However, in support of the modelling hypotheses for those systems not directly proven experimentally, modern science offers the powerful means of numerical simulations, whose results have to be compared with the reality that emerges from the experiments. The work of Rahman and Stillinger [57] is a shining example: they were able to predict the presence of a crossover between two regimes of propagation of elastic waves in liquids, in turn macroscopic manifestations of different microscopic situations, the ordinary sound and the second sound. Subsequent measurements confirmed this numerical modelling prediction [20,22-23,31].

Because of the increasing interest of the scientific community on Dual models of liquids, there are currently many published works on numerical modelling of liquids aimed at evaluating not only the validity of such models, but also the consequences of the duality of liquids on observable quantities. We will then refer on the results of such numerical modelling and quote their main results. In this frame the results achieved by Ghandili et al. [87] are extremely interesting. They have developed a numerical model to simulate a system arranged by means of molecular clusters, whose molecules may experience only quasi-harmonic vibrations around equilibrium position within the lattice, exactly as supposed in the DML, and free particles, i.e. those experiencing diffusive jumps. The authors approach the substance as a fractal lattice and introduce a new concept of thermodynamic dimension, $D_T$, corresponding to the average number of intermolecular interactions per *liquid particle* as function of T; it turns out to be a measure of how many molecules are bound in the lattice of a single *liquid particle*. The authors make use of the Maxwell-Boltzmann distribution function to describe the total fractional population of particles in the $n_{th}$ vibrational level, ad show that their approach may distinguish between free particles and bound particles defined above. Besides the model allows to introduce the value $\Theta_{max}$ for the liquid temperature corresponding to the border at which all DoF begin to be excited. For temperature below $\Theta_{max}$, the effective DoF for every direction is 1. One may see that $\Theta_{max}$ is directly linked to the parameter $m$ introduced in the DML. Accordingly to the above, at temperatures below or around $\Theta_{max}$, the system is characterized by $D_T = 3$, while this parameter tends to zero as the temperature increases (analogously to $m$). One of the first results is that $D_T$ allows to generalize Einstein and Debye models so that both are a special case of the general model. Then the authors propose a general expression for the vibrational DoF contribution to the isochoric specific heat (in a similar way as we have done in 76). They finally show that in a wide range of temperature and pressure, the calculated results agree very well with experimental data for isochoric heat capacities in dense region in the considered fluids. In particular, by introducing a new condition ($D_T = 1/2$) to plot the Frenkel line, they predict solid-like features around the critical point.

The same relevance, if not larger, has the work by Zhao et al [88]. The authors use the expression for $C_V$ deduced in the PLT model [61-65,67-72] to obtain an expression for the thermal conductivity of liquids, $K_L$. Within the same frame of reference of PLT and DML, i.e. by assuming a Dual nature of liquids, constituted by a mixed population of particles and phonons, they develop a method to calculate $K_L$. The agreement with experimental data is the highest ever obtained, even if compared with previous models of liquids.

Classical molecular dynamics simulation allows to study the time evolution of the system using Newtonian mechanics or its variants (Lagrangian or Hamiltonian equations of motion) and to model intermolecular potentials. In some cases it is necessary to resort to quantum molecular dynamics simulation of the system. A historical and general review of the main results obtained with these methods are collected in the Chen' review [7], while recently Proctor [89] has constructed a new numerical method to calculate, within the frame of the PLT Dual Model, the internal energy and heat capacity of liquids over a wide pressure and temperature range. Proctor has applied his numerical code to a number of real liquids in both the subcritical and the supercritical regime, in which the Frenkel line represents the border for the liquid state, in agreement with the Dual models. Experimental data have been fitted in a wide pressure–temperature range allowing to test the theoretical model with unprecedented rigor. The degree of accuracy of the prediction of internal energy and heat capacity is constrained by the different input parameters, such as the relaxation time of the liquid (initially obtained from the viscosity), the Debye wave number and the modulus of infinite frequency cut-off G. The model is successfully applied to deduce the internal energy and heat capacity for different fluids (Ar, Ne, $N_2$, and Kr) over a wide range of densities and temperatures. The author finds that the predicted heat capacities are extremely sensitive to the values used for the liquid relaxation time. One of the test beds for a liquid state model is indeed the unconventional trend of the specific heat of liquids with respect to temperature. In fact, if the specific heat of gases and solids remains almost constant with the temperature, or at most increases, the opposite happens in liquids, going from 3$\mathcal{R}$ at low temperatures, up to 2$\mathcal{R}$ at the critical point. The physical interpretation of the behaviour of specific heat in gases and solids, as is known, foresees a distribution of the thermal energy on the available DoF (energy equipartition principle): since the number of DoF increases in a substance with increasing temperature, this physical model is able to explain the observed phenomenology. Consequently it appears evident that this mechanism cannot be taken as the basis of a model of the liquid state, or at least it cannot be the only mechanism working at a microscopic level in a liquid. Dual liquid models, such as DML, answer to this need with the hypothesis that solid-like structures survive in liquids and participate in the distribution of available energy by means of their collective DoF. This idea, born with Brillouin and Frenkel, is today supported by the experimental evidence of the ability of liquids to effectively support and transmit high frequency transverse elastic waves, a key feature of solids [19-37]. The pseudo-crystalline structures, the icebergs, are dispersed in the amorphous liquid medium and contribute to the energy balance and to the propagation of thermal and elastic energy through their collective DoF, indeed they constitute the dominant part of it at low temperatures (and high pressures). However, since the number and extent of these structures decrease with increasing temperature, and the same do their collective DoF, their contribution progressively decreases in the energy balance of liquids, determining the observed trend of specific heat, as well as of sound propagation velocity (PSD). The expression for $C_V$ achieved in the DML (see Eq.20) allows to correctly reproduce such phenomenology from the theoretical point of view, being indeed strictly related to the free parameter $m$, the statistical number of collective DoF, that, by definition, decreases with temperature.

Even if Proctor's work is based on the PLT, what we want to highlight here is that substantially the results of the numerical simulation allow us to state that the Dual Models, as well as the DML, are suitable for describing the liquid state with sufficient accuracy. The investigations described by Proctor constitute the most detailed and rigorous test of the phonon theory of liquid thermodynamics to date. From the studies presented, the author concludes that the theory can model the internal energy of fluids over a wide (P,T) range with sufficient accuracy to model the heat capacity also. The unique behaviour of dense fluid heat capacity is successfully accounted for. Where input parameters are required other than the experimental data, one can confidently states that the parameters take the best-fit values that are physically reasonable, consistent with the available experimental data, and exhibit the expected trends as a function of temperature and density. The author comments that some discrepancies are reasonable because the PLT is an approximate treatment (unavoidably so since it is based on the Debye model for solids, which is itself not exact, a circumstance bypassed by DML thanks to the introduction of the parameter $m$).

For a model to have a physical meaning, it must be a model that may not fit the experimental data if it were wrong, as opposed to a model that automatically fits the data due to the number of adjustable parameters. The PLT has a sufficient number of adjustable parameters such that it can always be made to fit the data because very few direct measurements of Frenkel's liquid relaxation time are available, and these measurements do not cover a wide (P,T) range. However, what Proctor shows is that the PLT could have failed in a multitude of ways had it not been correct, but that was not the case. Said in other words, the theory is falsifiable but he wasn't able to falsify it. Proctor undertakes the same test in his recent book "Liquid and supercritical states of matter" [90].

Specific numerical simulation of the DML dealt with in this paper are currently aimed at evaluating the variation of the parameter $m$ with temperature, pressure, chemical species, etc to fit the experimental values of the specific heat, thermal conductivity, and of the other physical quantities. The values of $m$ thus obtained will represent the number of collective DoF participating in the microscopic dynamics of the liquid state [Peluso, F. et al., in preparation].

## 5. Discussion

Aim of this work is to illustrate the physical model dubbed DML which assumes that liquid molecules are arranged at mesoscopic level on solid-like local lattices. These clusters, or icebergs, are ephemeral dynamic objects, in continuous re-arrangement. Within these local domains of coherence, propagation of perturbations occurs at characteristic timescales typical of solids, while the local clusters interact each other by means of inelastic wave-packets exchanging with them energy and momentum. The interaction mechanism proposed is symmetric upon time reversal and the interaction term $f^{th}$ is deduced on the basis of "first principles".

The OoM calculation of the relaxation times, in particular of $\langle \tau_p \rangle$, is made possible by looking intimately on how a *lattice particle*, i.e. a wave packet, and a *liquid particle*, i.e. a dynamic iceberg, interact, as shown in Figure 3 and Figure 4. Calculations based on a kinetic approach show that in the case of water at ambient conditions the liquid particle is in full viscoelastic regime, with a prevalence to the single particle elastic regime by increasing the exciting frequency, as foreseen in the PLT and evidenced by experiments. In this conditions the particle may exchange energy and momentum with all the 3 types of DoF, 2 transversal and 1 longitudinal. This is due to the presence of the solid-like structures in which energy propagates at the speed $\langle u^0 \rangle$ as in solids, i.e. with reduced dispersion (harmonic waves). When the perturbation reaches or leaves a solid-like structure, anharmonic interactions arise travelling at the speed $\langle u^{wp} \rangle$, giving origin to a cut-off in the speed of sound and determining the transport of energy and momentum, allowing the structure to displace inside the liquid, alike a self-diffusion motion. This arrangement of liquids allows to justify the experimental values found on mesoscopic scale for the speed of sound close to that of the corresponding solid form, the characteristic trend of the specific heat vs temperature, and to account for the PSD with frequencies. When the speed of sound is measured at short wavelengths, i.e. on distances comparable to the extension of a local domain, a solid-like value $\langle u^0 \rangle$ is measured. By increasing the distance over which the sound speed is measured, hence on hydrodynamic scale, the classical value is found because it is averaged over distances much larger than the size of the a single cluster, involving the propagation through the liquid amorphous phase.

At equilibrium the flux of wave-packets is driven by the difference of elastic radiation pressure (Eq.(11) or (12)), that manifests through a local virtual temperature gradient $\left\langle \frac{\delta T}{\delta z} \right\rangle$ working as a generalized thermodynamic force. This is equivalent to assume a Local Thermodynamic Non Equilibrium determined by the heat current, although isotropically distributed in the liquid. The Non Equilibrium character is however related only over distances of the order of the wave-packet mean free path $\left\langle \Lambda_{wp} \right\rangle$. When the symmetry of the system is broken, for instance by imposing a temperature, a concentration or a pressure gradient, the correlation lengths increase up to some millimetres, according to the intensity and orientation of the gradients [48-50] (and to other experimental constraints). In the DML the presence of a gradient induces a preferential direction in the orientation and propagation of local, virtual heat currents (the flux of wave-packets). This flux is responsible for mass and energy transport in liquids out of equilibrium.

The elementary scattering events shown in Figure 3 cope with the presence of the two symmetric Brillouin lines in scattering experiments performed in liquids. The central line of a Brillouin spectrum is at the same frequency of the impinging beam and hence collects the energy due to elastic, or quasi elastic, scattering of the photons of the impinging light beam with the liquid' phonons constituting the heat current [3,11-13,53,91]. The two Brillouin lines are instead shifted with respect to the central line by a frequency proportional to the speed of sound in the liquid [Eq.(1)], therefore they account for inelastic scattering. At equilibrium, the two Brillouin lines are perfectly symmetric, because the amount of energy removed from the phonon' pool (Stokes line) by the impinging light beam is the same given back to the phonon' pool (anti-Stokes line). This is justified in the DML by the fact that the events a) and b) of Figure 3 are equally likely to occur (microscopic time reversibility). This last point recalls the attention of the reader on a very interesting aspect related to this elementary interaction mechanism at the base of the DML. The symmetry of this interaction upon time reversal can be related to the Onsager reciprocity relations [92-94], that require the existence of linear universal laws for molecular fluctuations to justify the time reversibility. The elementary interaction mechanism shown in Figure 3 and the non equilibrium statistics of *wave packet-liquid particle* interactions represent the physical base of the Onsager Principle.

When a temperature gradient is imposed to the liquid, the two Brillouin lines do not have anymore the same intensity [48-50,53 and Eqs. (1-3)]. To an external temperature gradient $\nabla T$ corresponds a heat flux (in the opposite direction), and therefore an increase of "*wave-packet ↔ liquid particle*" collisions at a given $\hat{q}$ vector; the intensity of the line accounting for this increase along the direction of the heat flux will be greater than the symmetric one by an amount proportional to $1/q^2$ (see Eq. (3)). In this case, events of type a) have a larger probability to occur than events of type b). Of course, if the symmetry is broken by a particle concentration gradient instead of a temperature gradient, the events having a larger probability of occurrence will be the b). To quantify such reasoning, let $\frac{dT}{dz} = \frac{T_1 - T_2}{L}$ be the temperature gradient externally applied to a system. As before, we assume here as first approximation that the application of an external temperature gradient does not change the average total number $\mathcal{N}^{wp}$ of the wave-packets, (this assumption is no more true in case large gradients are applied, because of non-linear effects that could arise). Let $\langle v_p \rangle$ be the average number per second of "*wave-packet ↔ liquid particle*" collisions at $T_0 \equiv \frac{T_2 + T_1}{2}$ in the isothermal liquid; due to the presence of the virtual temperature gradient, which is equally present in all the three directions, for each direction this number becomes $\langle v_p \rangle / 6$. If an external (small) temperature gradient $\frac{dT}{dz}$ is applied to the liquid, the total number of collisions per second $\langle v_p \rangle / 6$ is assumed to stay constant because we have supposed that $\mathcal{N}^{wp}$ does not change, but for collisions occurring in the direction of heat propagation there will be $\delta \langle v_p \rangle$ excess of collisions per second, and an equal defect $-\delta \langle v_p \rangle$ in the opposite direction. If the intensity of the temperature gradient is not too much high, it is reasonable to assume that there will be no non-linear effects and that the initial state of the system at the microscopic level locally continues to be similar to the one at uniform temperature, except for thermal excitations along z. Therefore a *liquid particle* at a given place in the liquid experiences the same number of collisions per second as if the temperature was uniform; the only difference with the isothermal case is that there is an imbalance in the number of collisions with wave-packets that are originated in the two half-spaces along z, one hotter and the other cooler. In the linear range we may assume that $2 \cdot \delta \langle v_p \rangle$ is proportional to the temperature gradient $\frac{dT}{dz}$ externally applied to the liquid. Consequently, there is a heat flux $J_q^{ext} = -K_l \frac{dT}{dz}$ superimposed to two equal and opposite heat fluxes $+ j_{+z}^{wp} = -K^{wp} \left\langle \frac{\delta T}{\delta z} \right\rangle_+$ and $j_{-z}^{wp} = -K^{wp} \left\langle \frac{\delta T}{\delta z} \right\rangle_-$ giving:

45. $\quad \frac{2 \cdot \langle \delta v_p \rangle}{2 \cdot \langle v_p \rangle / 6} = \frac{dT/dz}{\langle \delta T/\delta z \rangle}$.

The rationale of Eq. (45) is that macroscopic and microscopic (or mesoscopic) heat conduction obey the same phenomenological relations with identical coefficients [92-93]. Accordingly, an external temperature gradient $\frac{dT}{dz}$ increases by $\frac{dT/dz}{\langle\delta T/\delta z\rangle}$ the flux of phonons with respect to that corresponding to the local microscopic heat currents $j^{wp}$ due to spontaneous phonon diffusivity. Eq.(45) gives the ratio of the number of "*wave-packet ↔ liquid particle*" collisions per second due to the applied gradient, to that due to the random motions of collective thermal excitations. The clusters therefore will experience $2\delta\langle v_p\rangle$ collisions in excess along the direction of the heat flux and will execute as many jumps per second of average length $\langle\Lambda\rangle$ in excess in the same direction. Consequently, every particle travels the distance $2\langle\Lambda\rangle\delta\langle v_p\rangle$ per second along the direction of heat flux generated by the external temperature gradient. This quantity represents the drift velocity $\langle v_p^{th}\rangle$ of the *liquid particle* along $z$ due to the external temperature gradient:

46. $\quad \langle v_p^{th}\rangle = 2\langle\Lambda\rangle\langle\delta v_p\rangle = \frac{\langle\Lambda\rangle\langle v_p\rangle}{3}\frac{dT/dz}{\langle\delta T/\delta z\rangle}$,

and is the counterpart of the *wave packet* propagation velocity $\langle v_s\rangle$ defined in Eq.(8).

The above argument can be used in principle, and with the obvious adaptations, also for solute particles dissolved in liquid solutions to which a temperature gradient is applied.

Let's now come to Eq.(20) giving the expression of $C_V$. In a separate paper [76] it has been compared with the corresponding expression given by PLT model [61-65,67-72]. The Authors of PLT calculate an expression for the harmonic and anharmonic contributions to the specific heat within a liquid phonon theory based on the concept of liquid relaxation time introduced by Maxwell [8]. This approach, confirmed by measurements performed in 21 different liquids, covers the cases of gaseous, liquid and solid states of matter, giving for $C_V$ a theoretical limit value of $3\mathcal{R}$ for solids down to $2\mathcal{R}$ for liquids. Eq. (20) does not allow in the present form a direct evaluation of its dependence upon temperature (or other parameters), and in particular of the Arrhenius-dependence, $C_V \propto T^{-1}$. Nevertheless, the presence of $m$, or what is the same of $m^*$, and of $\frac{dm}{dT}$ allows us to speculate about the physical explanation of such a dependence. As pointed out above, $m$ accounts for the collective DoF actually participating to the energy distribution within a liquid. In a solid, the increase of $C_V$ upon temperature ($C_V \propto T^3$ Debye law) is due to the fact that the number of DoF that become excited increases with T. In the case of liquids, if we accept the picture of the DML, the mesoscopic structure is a mix of solid clusters and amorphous phase. The fact that the number and size of clusters decreases with temperature is reflected by the temperature dependence of $m$ that, in turn, provides a decrease of $C_V$ upon temperature. Very recently Zaccone and Baggioli [95] proposed a universal law for the vibrational density of states in liquids (VDOS) $g(\omega)$ using the Instantaneous Normal Modes (INM). This model has allowed for the first time to get an expression for the VDOS able to reproduce the functional dependence $g(\omega) \propto \omega$ typical of liquids at low energy, confirmed experimentally in several liquids [96]. Besides, the obvious consequence is that of providing the correct functional dependence $C_V \propto T^{-1}$ in liquids [97]. It is the case to underline that the Debye model $g(\omega) \propto \omega^2$ is not formally used in the DML, neither any of the expressions obtained for $C_V$ or any other thermodynamic parameter depend formally upon a specific distribution function for $g(\omega)$. Rather, it could be very interesting to speculate about some common features among the Zaccone-Baggioli model and the DML. A first question one may rise come from the comparison of the expressions of $C_V$: is there a relation between $g(\omega)$, as got from [95], and the parameter $m$, or $m^*$, introduced into the DML?. Again, where and how enter the INMs in the DML? INMs are by definition overdamped oscillatory modes characterized by having only the *Im* part. Being overdamped, they disappear quite immediately after having been generated. Are the INMs those modes that originate through the elementary *phonon-liquid particle* interaction? They indeed disappear quite immediately, lasting $\langle \tau_p \rangle$, of the order of few picoseconds (see Table 1). It is the case to underline that the value of the phononic contribution from Eq.(20) is always lesser than that of the total specific heat, as indicated in Eq.(23), the equality being valid in our hypothesis at low temperatures, where $m \approx 1$ and $\frac{dm}{dT} \approx 0$, and the phonons contribute about 100% to the total specific heat. On the contrary, the phononic contribution disappears at high temperatures, where there are no more collective excitations in the liquid, and the total specific heat decreases to lower values. The above behaviour is due again to the variation vs temperature of $m$, $\frac{dm}{dT} < 0$ [76].

PLT model has been verified so far on simple liquids, which show decreasing specific heat with temperature. However, specific heat for more complicated liquids can increase with temperature. It is believed [7] that the reason for this increase is the increasing contributions of internal vibrational modes of the molecules. In simple molecules such as water, the intramolecular modes vibrate at high frequencies and normally are not excited; consequently they can be neglected, as assumed by Trachenko and co-workers. However, with increasing molecular size, internal vibration frequency spreads to lower values, some of them can be thermally excited, and their number increases with increasing temperature and also their contribution to the specific heat as consequence. This contribution should increase with temperature because even more internal vibrational modes can be excited as temperature increases. Such vibrational contributions had been mostly neglected in thermodynamic analysis of liquids. Here in the DML this problem is overcome because of the introduction of $m$, that accounts for the DoF actually excited. Numerical simulations suitably set up [Peluso, F., et al, in preparation] will provide the (statistical) values of the DoF which put in agreement the theory and the experiments.

Very interestingly, according to recent theoretical findings phonons may carry gravitational negative mass. In their paper [98] Esposito and co-workers show that sound waves, constituted by phonons, not only transport mass during their propagation through condensed media (in the DML phonons carry also momentum), but also that such associated mass is negative. Indeed, in a density gradient they move towards a medium with a lower density, where the impedance is higher and their speed is lower. Of course this theoretical result could have a huge impact on what said before about the phononic contribution to the total heat content of a liquid [Eq.(15) through Eq.(21) and Eq.(23)]. If on one side Eq. (23) remains valid in principle, on the other side the phononic contribution to the total pool of thermal (and elastic) energy makes the "classical" part even more prominent with respect to the (negative) part due to collective modes. Indeed, the phononic heat content provided by Eq. (20) turns out to be negative, because the inequality in Eq. (21) is no more valid (and consequently that in Eq. (22)). It is then clear that, if the DML, as well the considerations made by the many Authors about the experiments performed on liquid systems by means of Brillouin scattering and the theoretical forecasts on the negative mass contributed by collective excitations are all valid arguments, an important link is still missing between theory and experiments on liquid structure, at least as it concerns the interpretation of the negative mass contributed by phonons in the experimental results. A first elementary, though puzzling, question to be answered is whether the II Principle is valid for a system in thermal equilibrium, in which phonons may subtract heat to the pool when colliding with the *liquid particle*, just because of the presence of an external gravitational field, responsible for the establishment of a density gradient in the isothermal liquid, or whether a "Demon" is working to make more reliable those events that are neglected in the daily physics reality [99]. In other words, is the Clausius postulate applicable to an isothermal system in presence of a gravitational field? Is the presence of this field enough to allow phonons to subtract heat to an isothermal system? Why do phonons prefer the more difficult path than the easier one? What makes easier the path that is normally the most difficult?

Some considerations are in order also for Eq.(27). First, as for $C_V^{wp}$, $K^{wp}$ accounts only for the fraction of thermal energy transported by collective thermal (elastic) excitations of the liquid crystalline lattice driven by the local virtual temperature gradient $\left\langle \frac{\delta T}{\delta z} \right\rangle$. Whether this is a small or a large fraction of the total, or accounts for the whole thermal energy, depends on various circumstances. In a crystalline solid this fraction should represent the total heat transported; indeed for such media, thermal (and elastic) energy is carried by means of phonons. In a metallic solid this is no longer true, due to the contribution of free electrons. What is the situation in ordinary liquids is not yet perfectly clear. According to the model presented here, if a direct measurement of $K^{wp}$ is not feasible its relative importance with respect to $K_l$ will depend on the fraction of heat transported by collective excitations, i.e. on $m$ (Eq.15). As it concerns the dependence on temperature of $K^{wp}$, it is mainly due to that of $u^{wp}$, and in turn to that of $\langle \Lambda_{wp} \rangle$, $\langle \tau_{wp} \rangle$ and, to a minor extent, to that of $\frac{dm}{dT}$. Evaluation of these quantities in some liquids may allow the calculation of $K^{wp}$ and of the involvement of collective oscillations in carrying heat in liquids. With the exception of water, that shows a behaviour strongly influenced by the hydrogen bonds, thermal conductivity in common liquids decreases with temperature, as expected also in the DML frame; this is at least due to the presence of the quantity in square brackets, that decreases with temperature (as also the other quantities in the Eq.(26) for $K^{wp}$). From the DML point of view, the decrease of $K^{wp}$ with temperature is due to the decrease of the number and size of the solid-like dynamic icebergs and, in turn, to the decrease of the available collective DoF. The presence of the terms in square brackets, in turn, ensures the positive definition of $K^{wp}$ through Eq.(20). Nevertheless, the considerations raised for Eq.(20) related to the negative contribution of the phononic part to the total pool of energy content of a liquid, are valid even for Eq.(26).

A final consideration on $K^{wp}$ is that the DML opens the perspective to a heat propagation in liquids modelled in terms of a Cattaneo equation. Indeed the delay-time requested by this type of equation is physically justified, representing the average time spent by heat in non-propagating forms (tunnel effect). Interestingly, it has been shown [100] that in a thermo-elastic model for liquids, in which Eq.(13) is assumed as a constitutive equation for the system, the consistency of the model with the II Principle is ensured only if a Cattaneo equation is assumed to hold for heat propagation. This argument is the subject of a separate paper [80].

Some remarks on the role of relaxing phenomena acting in such a system are also in order. Let's return to Eqs.(4) and (5) giving the energy and momentum exchanged between solid-like clusters of liquid molecules and wave-packets, and to Eqs.(6) and (7), for the time lapses and particle displacements involved in the collision process. As shown in Figure 3 and Figure 4, the energy lost by the wave-packet is commuted into kinetic (i.e. translational) and potential (i.e. collective vibrations) energies. Each of the two energy pools will be dissipated once the collision is completed, and a relaxation time can be defined for each process. The kinetic part of the energy of the *particle*, involving external DoF, is dissipated by means of friction with a own decay time $\tau_k$, according to the decay process

47. $\quad \Delta E_k^p(t) = \Delta E_k^p(t=0) \cdot e^{-t/\tau_k}$.

The part responsible for the excitation of the internal collective vibratory (quantized) DoF will be relaxed with relaxation time(s) $\tau_{\Psi_i}$ pertaining to the several excited distinct DoF, and following the decay process [101]

48. $\quad \Delta \Psi_i^p(t) = \Delta \Psi_i^p(t=0) \cdot e^{-t/\tau_{\Psi_i}}$.

As pointed out by many Authors [101-106], a distribution of relaxation times has been found in complex systems, as liquids, rather than a single relaxation time. However, being this specific topic outside the scope of the present paper, we still consider a single relaxation time $\langle \tau_R \rangle$ for our system because we are here only concerned with a comparison of the several time intervals introduced in our model with the wave-packet lifetime. In fact, depending on the length of the relaxation time $\langle \tau_R \rangle$, defined in Eqs.(7), (47), (48), with respect to the wave-packet lifetime $\langle \tau_{wp} \rangle$, defined in Eq.(16), the dynamic iceberg may interact with another phonon before or after $\langle \tau_R \rangle$ has elapsed [this depends also on the wave-packets density in liquids, see Eq. (39)].

Experimental investigations performed in several liquids have shown that the scattering spectra are characterized by two distinct branches: the so-called viscous, or hydrodynamic-like regime, and the elastic, or solid-like regime [19-37], as foreseen also by numerical simulations [29,57,87-90]. These branches may be explored in the sub-THz or THz frequency range, and in all cases the spectra (INS or IXS) show a PSD, i.e. an increasing value of the sound speed from low to high frequencies, or alternatively, from low to high wave-vectors; said in other words, PSD consists in a gradual increase of sound velocity upon increasing of the exchanged energy $\Delta \varepsilon$ or of the exchanged momentum $\Delta p$. This is shown in the two plots reported in Figure 5, where data extracted from [34] for water and from [19] for glycerol are reported. What immediately jumps to the eye is that the speed of sound spans from the hydrodynamic value, at low momenta, to the solid-like value, at high momenta. Incidentally, for water the solid-like value is that found for the first time by Ruocco et al in 1996 [21]. The propagation of acoustic, or elastic, waves is associated to the compression and rarefaction of the medium, or density fluctuations. At low frequencies, or low propagation speeds, the viscous regime is characterized by relaxation times much shorter that the period of oscillation of the wave packets $\tau = (\tau_p + \tau_R) \ll \tau_{wp} = 1/\nu^{wp}$, where $\nu^{wp}$ is the wave-packet frequency [19,34, see also Eqs. (6) and (16)]. In other words, the target sample relaxes so quickly that the acoustic perturbation propagates over successive equilibrium states. On the contrary, at high frequencies and high propagation speeds, the elastic regime is characterized by periods of oscillation of elastic perturbations much shorter than the relaxation times, $\tau_{wp} \ll \tau$. In this case the acoustic oscillations are too much quick to allow the DoF to relax their energy before a second perturbation hit the *liquid particle* again. In this regime the perturbations propagate elastically, with very much reduced energy loss. The intermediate branch, i.e. that in between the viscous and the elastic regimes, is usually referred to as the viscoelastic regime; it corresponds to the energy, or momentum range in which the liquid undergoes the transition from viscous to elastic response. This specific range corresponds to the occurrence of $\tau/\tau_{wp} \cong 1$, from which one may deduce the value of the relaxation time, $\tau \cong 1/\nu^{wp}$. The increase of momentum and energy exchanged between phonons and *liquid particles* in the viscous-to-elastic transition or, what is the same, the decrease of acoustic dissipation, brings us to conclude that at low frequencies, or low sound velocity, the mechanism at the base of the wave packets ↔ *liquid particles* interactions is mainly collisional and involves long wavelengths; at high propagation speeds, and therefore at high frequencies, the transmission-propagation mechanism of elastic waves involves also structural modifications, activating internal relaxation phenomena; in this case the interactions are characterized by short wavelengths. One may readily conclude that the coupling of a sound wave with a specific DoF involves not only a characteristic time-scale, through the relaxation time, but also a length-scale, through the exchanged momentum, $\langle \Lambda_p \rangle \approx \lambda = 2\pi/q$ (indeed, for a wave-packet ↔ *liquid-particle* interaction to occur, the wave-packet characteristic wavelength, or its extension, should be large at more as the *particle* size, if not smaller). Figure 5 shows us that a relaxation time due to internal DoF makes the sound velocity depend also on the exchanged momentum and hence on the probed distance [19,34].

The calculation of the *liquid particle* displacement and of the time interval between two successive interactions is not trivial, and only in very few special cases one is able to calculate their values, as for instance in the case in which all the energy (and momentum) lost by the wave-packet would be acquired by the *particle* and converted into kinetic energy, i.e. when only the translational DoF are involved in the scattering process. In the case dealt with here, the interaction is not instantaneous because it is anharmonic and the momentum exchanged during the interaction is proportional to the duration of the interaction itself, $\langle \tau_p \rangle$ (i.e. to the time-length of the wave-packet). The energy lost by a wave-packet in an event of type a) of Figure 3 is partially converted into kinetic energy of the *particle* and partially into internal molecular energy, exciting the corresponding DoF. This is an important point of the DML that is the first phonon model of liquid dealing with the exchange of energy with the internal vibratory DoF of the molecular cluster. Hertzfeld and Litovitz [101] introduced an *"effective heat capacity"* $C_{eff}$ as the total heat absorbed by, or given to, a *liquid particle*, considering the two branches of DoF. $C_{eff}$ is a complex quantity whose real part is the static part pertaining to external translational DoF, while the imaginary part depends on that pertaining to the internal vibratory (quantized) DoF, on the frequency and on the relaxation time. $\langle \tau_R \rangle$ is the time lag after which the energy stored into the cluster is released. Depending on its OoM, the events of Figure 3 can be considered elastic, quasi-elastic or inelastic [24,62,65,81]. Results of Section 3.4, already discussed at the beginning of the present Section, are in line with such argument, with the experimental outcomes and with the forecasts of PLT.

The comparison with recent developments in theoretical models on systems exhibiting $k$-gaps [102] provides very interesting insights. The gap in $k$-space in liquids can be related to a finite propagation length of shear waves; indeed the $k$-gap emerges in liquids only in the transverse spectrum, while the longitudinal one remains gapless. For the $k$-gap to emerge in the wave spectrum, two essential ingredients are mandatory. The first consists in getting a wave-like component enabling wave propagation; this is represented in the DML by the wave packets. The second is to get a dissipative effect that disrupts the wave continuity and dissipates it over a certain distance, thus destroying waves and giving origin to the gap in $k$-space. This last is represented by the wave packet – *liquid particle* interaction, that is the source of dissipation, and works displacing the wave packets from one place, where it is absorbed by the liquid particle, to another, where it returns to the system' energy pool, alike a tunnel effect. If $\langle \tau \rangle$ (Eq.(6)) is the time during which the shear stress relaxes, then $\langle \Lambda \rangle = \langle v_s \rangle \cdot \langle \tau \rangle$ (Eqs.(7) and (8)) gives the shear wave propagation length (or liquid elasticity length). Frenkel [5] already deduced a Cattaneo-like propagation equation, following an *ad hoc* phenomenological procedure he introduced into the Navier-Stokes equation the contribution due to the shear in a liquid (although he did not solve the equation he arrived at).

When the energy of the system does not change overall, the propagation range of a collective mode acquires a finite range; this induces an effect distinct of dissipation. Propagation of plane waves in crystals occurs without dissipation because the wave is an eigenstate, however plane waves dissipate in liquids because they are systems with structural and dynamical disorder. Particle dynamics in liquids is characterized by both oscillatory motions at quasi-equilibrium positions, allowing liquids to show a solid-like behaviour when experimentally investigated over mesoscopic scales, and diffusive jumps into neighbouring locations, enabling liquids to flow; jumps are in turn associated with viscosity [5,19,65,68,79,102]. This dynamics requires a non-linear interaction – as for instance the *liquid particle – lattice particle* interaction - allowing for both oscillation and jumps activated over potential barrier of the inter-particle potential. These are the typical interactions dealt with in Nonaffine treatment, that allows indeed to account for the Nonaffine displacements at atomic/molecular level. Liquids in fact do not have the simplifying features of solids and gases, i.e. small displacements and small interactions, while on the contrary they combine strong interactions with large displacements and for this reason cannot be described analogously to gases or solids but need of specific theoretical tools [6].

Baggioli et al. [102] used a two-fields potential representing displacements and velocities, $\phi_1$ and $\phi_2$, to build up the Lagrangian describing systems with **k**-gap constituted by two mutually interacting sub-systems (that we identify in the DML with the wave-packets and the *liquid particles*). Neglecting the details of the mathematical formalism, what matters is that the equations of motion for the two scalar fields decouple, leading to two separate Cattaneo-like equations for $\phi_1$ and $\phi_2$:

49. $$\begin{cases} \phi_1 = \phi_0 \exp\left(-\frac{t}{2\langle\tau\rangle}\right)\cos(kx-\omega t) \\ \phi_2 = \phi_0 \exp\left(\frac{t}{2\langle\tau\rangle}\right)\cos(kx-\omega t) \end{cases}$$

Here-to-follow we just cite the more relevant key-points emerging from the solution of the above equations for our purposes:

1) the interaction potential is an oscillating function (see Figure 5 in [102]), i.e. $\phi_1$ and $\phi_2$ reduce and grow over time $\langle\tau\rangle$, respectively; this implies that the two interacting sub-systems represented by the two scalar fields, exchange energy among them. Therefore we may hypothesize that the interaction between the population of wave packets and that of liquid particles is described by such couple of mutually interacting potentials.

2) because the total scalar field is the product of $\phi_1$ and $\phi_2$, the total energy of the whole system does not vary with time, i.e. it is a constant of motion, as expected in the DML in isothermal conditions for systems constituted by the two populations of mutually interacting sub-systems.

3) The motion (see Figure 5 in [102]) is a typical dissipative hydrodynamic motion. If the anharmonic interaction described by the Lagrangian, i.e. the wave packet – *liquid particle* interaction has a multi-well form, the field may not only oscillate in a single well, but also can oscillate from one minimum to another, giving origin to the hopping motion described in Figure 2. This motion is analogous to diffusive particle jumps in a liquid and may represent also a possible microscopic origin for the viscosity. The motion of the particles happens via thermal activation or tunnelling, between different wells with frequency $\nu_F = 1/\tau_F$. The dissipation varies as $\nu_F = 1/\tau_F$: large $\langle\tau\rangle$ corresponds to rare transitions of the field between different potential minima.

The presence of **k**-gaps in liquids may be associated with the fast sound or positive sound dispersion (PSD). Frenkel was the first to note that a non-zero shear modulus of liquids would have implied a cross-over of the sound propagation velocity from its hydrodynamic value $v = \sqrt{B/\rho}$ to the solid-like elastic value $v = \sqrt{(B+4/3\,G)/\rho}$, where B and G are bulk and shear moduli, respectively. Propagation of shear modes occurs at high **k** values, implying

PSD at these *k* points and further that PSD should disappear with temperature starting from small *k* because the *k*-gap increases with temperature.

Upon the revision process one of the Reviewer drew the attention to a general theory of collective phonon modes (dealt with as Goldstone bosons in crystal, liquids and glasses) recently presented [103]. This model, in which hydrodynamic and field theoretical methods are gathered, is based on Nonaffine displacements and the consequent phase relaxation of collective excitations. It has many interesting implications, among them the presence of shear waves above a critical value for wave-vector in liquids (*k*-gap), for the first time assessed on a theoretical firm basis, as well the theoretical explanation of PSD in liquids for longitudinal waves. One of the key-points of this model is the relaxation phase $\Omega$, that we identify as the inverse of the relaxation time $\langle \tau \rangle$ in the DML. $\Omega$, or $\langle \tau \rangle$, determines the propagation distance of the collective phonon modes through the typical exponential decay law $e^{-\Omega \cdot t} \equiv e^{-t/\langle \tau \rangle}$. The conclusion arrived at by the authors is that a liquid may be seen as a system in which the phonon collective modes are confined in clusters of size $\langle \Lambda \rangle \approx u^0 \cdot \langle \tau \rangle$. At larger distances the system looses the rigidity typical of solids and the associated capability of propagating the mechanical stress by means of shear waves. The reader will have already recognized that this is exactly the picture on which the DML is based, already described above and schematically represented in Figure 1. It is then very surprising, if not a unique circumstance, that two separate and independent groups have reached the same conclusion on the microscopic dynamics of liquids, although starting from different perspectives.

Summarizing, which is the dynamics we imagine in a liquid? A wave-packet travels at the speed $u^{wp} = \langle \Lambda_{wp} \rangle / \langle \tau_{wp} \rangle$ until the border of a lattice (Figure 1), where it interacts with the *liquid particle* as shown in Figure 3 and by Eqs. (4) and (5). Part of the energy lost by the wave packet is commuted into kinetic energy of the cluster, part excites internal quantized DoF (harmonic contribution). The perturbation inside the solid-like dynamic structure propagates by means of (quasi) harmonic waves, travelling at speed $u^0 = \langle \Lambda^0 \rangle / \langle \tau^0 \rangle$. The internal DoF will take a finite time to de-excite (Figure 4). As much longer will be this time with respect to the average lifetime $\langle \tau_0 \rangle$ of the elastic perturbation inside the local lattice, so long will be $\langle \Lambda_R \rangle$. The change of order due to the collision between the wave packet and the *liquid particle* contributes to the specific heat, the thermal conductivity and other liquid parameters. Such change of order requires energy to be added to, or taken from, the substance because of the change in potential energy associated with the change of structure. The time taken for such a change to occur is the relaxation time, and if the external change is too rapid, $\langle \tau \rangle \gg \langle \tau_{wp} \rangle$, the required structure change cannot follow it. Said in other words, the system takes too much time to lose memory of its past with respect to the time taken by another external stimulus to occur. The relaxation time increases rapidly with decreasing the temperature. At temperatures where the relaxation time amounts to about 30 minutes, the order cannot adjust itself to equilibrium during the usual time an experiment takes. The momentary state of order is "frozen in" and the material is in the "glassy" state, which exists below the glassy transition temperature. The glassy state parameters are more like those of a solid than those of a liquid (in the classical meaning). The glass has shear rigidity because the stress is applied over times much shorter than the shear relaxation. The experiments performed in the viscoelastic regime of liquids have shown that such behaviour persists in liquids for stresses of the duration of picoseconds. Therefore, in their behaviour in propagating ultrasonic waves, liquids may appear "glassy" at much higher temperatures (shorter relaxation times) than for ordinary experiments. They then behave like solids on short length scales, giving little absorption and showing, for transversal waves, shear rigidity. Short relaxation times, $\langle \tau \rangle << \langle \tau_{wp} \rangle$, indicate phenomena that relax before a *liquid particle* has undergone the next interaction. In this case the wave-packet sees in the next interaction a system that is modified (relaxed) with respect to the previous collision. In the opposite case, $\langle \tau \rangle >> \langle \tau_{wp} \rangle$, the wave-packet finds a system unmodified with respect to the former event, as in a rigid network [19,64].

Let's discuss how particle dynamics changes in response to pressure and temperature variations. In the DML particle dynamics can be separated into solid-like oscillatory and gas-like diffusive components. By increasing the temperature, each particle spends less time oscillating and more time jumping; by increasing the pressure the behaviour is reversed and results in the increase of time spent oscillating relative to jumping. Increasing temperature at constant pressure (or decreasing pressure at constant temperature) eventually results in the disappearance of the solid-like oscillatory motion of particles; all that remains is the diffusive gas-like motion. This disappearance represents the "qualitative" change in particle dynamics and gives the point on the Frenkel Line. Recalling indeed the historical definition of Frenkel line, it separates the low temperature liquid-like state, where particle dynamics combines with oscillatory and diffusive components of motion and where transverse modes exist, from the high-temperature gas-like state, where particle dynamics is purely diffusive and where no transverse modes operate, but only the longitudinal mode propagates. We could say that the Frenkel Line corresponds to the crossover from the ''rigid'' solid-like liquid to the ''non-rigid'' gas-like fluid.

The reader has certainly wondered whether it is possible to get an expression for the lattice potential that gives rise to the interaction term $f^{th}$. Even if a starting point could be the expressions given in Eqs. (49), an even more fundamental question is whether wave packet-*liquid particle* collisions can be treated as a classical scattering or it is necessary to examine them from a quantum point of view to analyze the consequences of energy absorption and re-emission, in particular for the internal collective vibratory DoF. Let's answer this question by considering the motion of a free particle, for instance a single electron. Its motion is not quantized in the external DoF, but only in the spin. However, if this particle is immersed in an external force field, then its motion will be quantized too, as for example an electron in the Coulombian field generated by a positive charge, as in the hydrogen atom. It will not be able to travel any orbits, but only those which satisfy certain conditions dictated by the rules of quantum mechanics. These conditions also limit the translational DoF of the electron; in fact the energy, the total angular momentum (modulus) and the azimuthal component are quantized. Of course, this does not exclude that an isolated atom can freely displace or rotate on itself. If, however, this atom is in turn immersed in a force field, for example because it is part of a molecule or of a larger lattice, then its global motion will also be quantized. His internal rotational and vibratory DoF will then also be quantized. This reasoning can be extended to even larger systems, at least within the limits imposed by the Heisenberg principle, which delimits the transition from the quantized world to the classical one. The statistical nature of the wave packet –*liquid particle* interactions does not allow in its current formulation a hypothesis about the lattice potential of a liquid, so that the two populations, *lattice particles* and *liquid particles*, which make up the liquid in the DML, are in fact supposed not interacting at a distance, but only when there is an overlap of the respective wave packets, from which the origin of the finite interaction time $\langle \tau_p \rangle$. The dynamic effect is generated only during the interaction itself, although the two subsystems are not a-priori brought to interact, like two particles driven by a potential. One may wonder if this fact could be an indication of the short-range character of the lattice potential. In the DML the "contact" character of the interaction leads to suppose the *liquid particles* as "free" before and after the interaction. Based on this hypotheses, it is plausible to assume the translational and rotational DoF of the entire *particle* as non-quantized (if the links with the other *liquid particles* are neglected). Therefore, only the internal vibratory DoF would remain quantized.

It is very interesting to compare the DML and the PLT models. The PLT cannot provide a right away answer on "why" the liquid particles oscillate because it describes the liquid from a statistical-thermodynamic point of view. DML instead starts from the microscopic elementary *wave packet* ↔ *liquid particle* collision that gives rise to the harmonic and anharmonic contributions. Thus it faces the problem from the alternative mesoscopic point of view. DML provides and analyzes the elementary mechanism by which energy and momentum are exchanged between *liquid particles* (icebergs) and *lattice particles* (phonons) and shows that this scenario is in agreement with that of the PLT. The relaxation time is one amongst the junction points between the PLT and the DML. PLT, because of its thermodynamic-statistical nature may only provide an indication of the variation range of the relaxation time by means of the value of the product $\nu\tau$. The mesoscopic character of the DML allows instead investigating the intimate mechanism of interaction of liquid particles with lattice modes; in this way DML provides both the answer to the question of "why" liquid particles oscillate, and a way for calculating the order of magnitude of $\tau$ in ordinary liquids.

We have already shown that the values of $\tau$ and of the product $\nu\tau$ are in the range foreseen by PLT. The relaxation time is one among the key-points of both models, and because the only shared argument of PLT and DML is the assumption that thermal energy in liquids is transported by means of collective lattice excitations, both harmonic and anharmonic, the mutual agreement shows that both of them look at the model of liquid state in the same way, although from different observation points.

PLT and DML share also the presence of harmonic and anharmonic contributions to the energy pool. PLT takes this into account by considering both contributions into the Hamiltonian of a liquid, resulting in turn in the expression for the specific heat. On the other side, harmonic and anharmonic oscillations are believed in the DML as the way for energy and momentum to propagate within a liquid. In fact, energy is supposed to propagate inside the ephemeral icebergs by means of harmonic oscillations. *Liquid particles* communicate among them and with the disordered liquid by means of anharmonic wave-packets, enabling the exchange of energy and momentum. In such a way the phenomenological effects of diffusion (and thermal diffusion) may be easily interpreted in the DML.

As we have seen in [76], the duality of liquids does not exclude that classical mechanisms of molecular interactions can work simultaneously at the mesoscopic level with interactions between wave packets and pseudo-crystals, both allowing propagation of energy. The system' thermodynamic conditions will let one of the two prevail on the other. The solid and gaseous states are one the opposite of the other, in the first only the phononic part will be present, while in the second energy propagates only by means of intermolecular collisions. At the melting the solid molecules begin to arrange in local networks and the phononic contribution gradually leaves way for energy propagation through molecular interactions, until they completely disappear when the gaseous state is reached. While in solids there are always three modes of vibration, in liquids the two transverse modes will participate to the energy distribution only for frequencies higher than $\omega_F$, $\omega > \omega_F = \dfrac{2\pi}{\tau_F}$. This view in the DML agrees with the interpretation underlying the PSD (see Figure 5). The propagation speed of elastic and thermal waves in liquids reaches values close to those in the corresponding solids as the frequency increases. On a macroscopic scale, at low frequencies, i.e. large wavelengths, it is not possible to reveal the presence of the local lattices, while they show their presence when the wavelengths become comparable to, or smaller than, the size of such clusters. As consequence of the above, the number $m$ of lattice collective DoF available in a liquid, introduced in Eq.(15), decreases as the temperature increases, $dm/dT < 0$. At low temperature, when the two transversal modes and the longitudinal one are excited, $m$ approaches unity, while it approaches to zero at high temperatures because the number of lattice collective DoF decreases with increasing T, $0 \leq m \leq 1$, until they completely disappear at the Frenkel line. On the other hand, $C_V$ is a physical quantity very sensitive to the phase transition, and its continuity across the triple point could be attributed to a continuity also of the solid and liquid mesoscopic structure. This is a fundamental point of the DML. In [76] this argument, as well how the relative importance of intermolecular collisions vs collective wave packet – *liquid particle* collisions affect the isochoric specific heat of liquids, are deeply discussed and limits for $dm/dT$ and $m$ are deduced and analysed.

The evaluation of the crucial parameters rests then on that of $\langle \tau_0 \rangle$, $\langle \tau_p \rangle$ and $\langle \tau_R \rangle$, and hence on the physical modelling of the relaxation processes involving internal DoF [24,65,81]. An interesting theory has been proposed by Nettleton [104-106] that introduced a set of phenomenological equations, valid at all frequencies, for the rates of change of internal relaxation parameters describing molecular excitations and structural changes in a liquid. He found an Onsager coupling between the rate equations and the pressure tensor.

The values of phonon mean-free-path $\langle \Lambda_0 \rangle$ and life-time $\langle \tau_0 \rangle$ within a *liquid particle* can be deduced from the experimental values obtained in light scattering experiments. Indeed, $\langle \Lambda_0 \rangle$ will be a multiple of the phonon wavelength $\lambda^0$, $\langle \Lambda_0 \rangle = n\lambda^0$, and $\langle \tau_0 \rangle$ of $\tau = 1/\nu^0$, $\langle \tau_0 \rangle = n/\nu^0$, with $n > 1$. Using the data for water of [19,34], typical values for the parameters characterizing a phonon (variation range is function of temperature, pressure and $q$ orientation) are: (central) frequency $\frac{n}{\langle \tau_0 \rangle} = \langle \nu^0 \rangle \approx 0{,}95 \div 2{,}5 \, THz$, wave-length $\frac{\langle \Lambda_0 \rangle}{n} = \langle \lambda^0 \rangle \approx 1 \div 3 \, nm$ and velocity $\frac{\langle \Lambda_0 \rangle}{\langle \tau_0 \rangle} = \langle \lambda^0 \rangle \cdot \langle \nu^0 \rangle = \langle u^0 \rangle \approx 3.1 \div 3.4 \cdot 10^3 \, m/s$. Interestingly, this value fits very well the experimental data obtained for the propagation velocity of thermal waves in water [21]. The value of $\langle \Lambda_0 \rangle$ thus obtained represents also the typical OoM of the size of a *liquid particle* at the exchanged momentum $q$. We may then confidentially assume a value of few units for the parameter $n$.

Regarding the possibility of thinking to the external temperature gradient as the generalized force, the energy flow generated by it is modelled as a variation of the wave-packet energy, having neglected so far any variation in their density. The effect due to the local, or virtual, temperature gradient is different, it could be responsible for a local curvature of the lattice, which in turn is responsible for the phonon-*liquid particle* interaction. This is an interesting topic that will be further dealt with in a future work.

## 6. Conclusions

In the Dual Model of Liquids the inertial term $f^{th}$ provides a physical mechanism by which a cluster of molecules and a liquid lattice exchange energy and momentum. The elementary interaction at the base of this process and how it works are shown in Figure 3 and Figure 4. Its anharmonic character allows momentum exchange, and has the important property of being temporally reversible, as the non-equilibrium thermodynamics requires. In this frame, the experimental results obtained in collective dynamics of liquids by means of INS and IXS experiments may be physically interpreted. The model is also able to postdict the physical interpretation of some parameters typical of liquids and of other well known phenomena. DML provides also a comparison with some results obtained by independent research groups on the dynamics and thermodynamics of liquid media (liquid specific heat, thermal conductivity, sound velocity, capability of transporting momentum, etc.). In [76] a direct comparison is made between the expressions for the specific heat arrived at in the DML and PLT. This comparison provides interestingly the minimum and maximum numerical values of the number of collective DoF involved in the exchange of energy and momentum, $m$, and of its derivative with respect to temperature, $\frac{dm}{dT}$.

The elementary interaction between *liquid particles* and *lattice particles*, allowing exchange of energy and momentum between the two pools, is at the base of a possible model for the liquid viscosity, and hence, it provides also a mechanism by which account for the dissipation [107]. In a physical model for viscosity of liquids the exchange of momentum and energy among the two subsystems should be accounted for, for instance by a suitable modification of the Navier-Stokes equation by using a two-fields Lagrangian. Another interesting topic to be investigated is the transition from inviscid liquid to gas (although a speculative explanation has been already advanced in the Discussion). Regarding this topic it is relevant to compare the DML framework with that emerging from the nonaffine framework [108], because the redistribution of energy between internal and external degrees of freedom of the clusters is reminiscent of a microscopic description of liquids and amorphous solids based on nonaffine displacements/deformations at the atomic/molecular level (see also Figure 2). Nonaffine treatment has been used to successfully explain the dependence of viscoelastic properties of liquids upon mesoscopic confinement size [109-114]. Experiments performed by Noirez and co-workers revealed that liquids confined to sub-millimeter scale show an elastic response to mechanical solicitations; in particular, their response to low-frequency mechanical stress is typical of amorphous solids, exhibiting a storage shear modulus $G'$ much higher than the loss viscous modulus $G''$. By increasing the frequency of the mechanical stress and/or the scale on which such stress is applied, the effect disappears, and the liquid returns to its classical full-viscous behaviour ($G' = 0$). Nonaffine framework has provided the physical interpretation to such behaviour, consisting in the cut-off of low frequency transverse modes, that in turn allows the high frequency modes to maximize their effect at mesoscopic level. In this framework it is assumed that transverse modes are carried by phonons. In this scheme the DML may provide the missing ingredient to nonaffine treatment by means of the inertial term $f^{th}$ responsible for the *phonon – liquid particle* interactions (Eqs. 13 or 12), as the best candidate for the nonaffine force introduced in the equation of motion of the *microscopic building block* [114]. In fact, in the nonaffine framework there is a competition between rigid-like behaviour of coordinated clusters and the relaxations via nonaffine motions due to unbalanced forces in the disordered molecular environment.

Subsequent investigations of the experiments described above allowed also to reveal an "unexpected" thermo-elastic effect in liquids confined at the mesoscopic level [115-117]. The effect is highlighted in two types of experiments, whose difference consists in the motion law of the movable disk. In both cases the mechano-thermal effect consists in the occurrence of a temperature gradient inside the liquid due to the momentum transferred by the moving plate to the confined liquid. In one type of experiment, the movable plate, typically the lower one, undergoes a sudden acceleration in one direction, performs the expected movement and then stops for a few seconds. Then it goes back to the starting point but with an acceleration lower than that of the outward path. In the second type, the moving plate oscillates according to a periodic motion. The thermal behaviour of the liquid in the two cases is not exactly the same, although the dependence of the intensity of the mechano-thermal phenomenon, in particular of the temperature difference that originates with respect to the distance between the two plates, remains almost the same. What matters here is that DML can provide, at least qualitatively, a physical explanation for this "unexpected" phenomenon. The key is provided indeed by the elementary phonon – liquid particle interaction, described in Figure 3. The acceleration of the moving plate generates a momentum flux, directed towards the fixed plate. According to the DML, and also following Rayleigh's reasoning, the momentum flux, dimensionally a pressure, generates a temperature gradient. The "liquid particles", pushed by the momentum flux, transfer their excess impulse to the phonons of the liquid, which emerge from each single impact with an increased energy. They are then pushed and collected towards the fixed plate, giving rise to a temperature gradient directed towards the fixed plate. The phenomenon is very rapid because it is determined by a succession of (almost) elastic collisions, which therefore propagate very quickly. Obviously, the higher thermal content of the liquid in proximity to the fixed plate compensates for the lower thermal content in proximity to the movable plate, leaving the thermal balance unchanged. It is here important to highlight that the phonon – liquid particle collision also explains why a temperature gradient (i.e. a generalized flux-force crossed effect) is generated inside the liquid instead of a simple heating due to the energy stored as consequence of the mechanical stress $(G' \gg G'')$, as one would have expected on the basis of a classical interpretation of the elastic modulus[7].

When the plate stops, the system relaxes and reaches the temperature it had at the beginning. When the movable plate returns to the starting position, the liquid is in the exact same condition it had at the beginning of the experiment, and the temperature gradient forms again. This interpretation is in line with all the basic hypotheses made by the Authors, including the one on the number of Grüneisen. In this particular case the relationship should not be read as an expansion/compression effect of the material medium, but rather of the density of the phononic contribution. Major details will be provided in a dedicated paper [Peluso, F., et al., in preparation]; in the meanwhile some suggestions/questions for further experimental set-up are advanced [118], namely, 1. It would be very interesting if it were possible to carry out a scattering experiment in the liquid cavity, although this would imply serious experimental difficulties; 2. To carry out the experiment by moving both plates in opposite directions (provided to have a suitable rheometer). If, on one side, this certainly increases the momentum flux transferred to the liquid, and hence to the phonons, on the other it becomes extremely interesting and instructive to see "if" and "how" the temperature gradient is formed, and also verify the $L^{-3}$ rule. 3. To evaluate a possible effect due to the extension of surface area of the two plates. 4. Perform the experiment with continuously rotating plate(s) One could expect that in this configuration the temperature gradient should survive for a longer time before relaxing spontaneously, as already demonstrated for the shear strain effect, [110].

One of the peculiarities of the interactions between wave packets and *liquid particles* is to work as a tunnel effect, by moving specific quantities of energy from one place to another in a well-defined time interval, the relaxation time, during which such energy packets are kept out of the heat current. The role of relaxation times is important in all the situations in which transport phenomena are influenced by a delay occurring in the microscopic transport processes. One of these cases is that of the heat transport in non isothermal conditions, where such delays have a relevant role in the correct identification of the physical process and of the related mathematical equations describing the heat propagation in condensed media. It is well known indeed that in non isothermal conditions a Cattaneo-like hyperbolic equation should be used to describe the heat propagation instead of a Fourier-like parabolic one to describe the heat diffusion. In [80] it is shown how the DML provides the physical interpretation of the delay time appearing in the Cattaneo equation.

---

[7] Given its small value, the temperature gradient is completely hidden when there are large volumes/layers of liquid. Furthermore, since the effect is due to a velocity gradient and not to a mere difference, in order to reveal it on a large scale, strong angular

There are many other insights deriving from the model itself and from its dual character. One of these deals with the PSD. Although a PSD was reported in a large class of fluids, it is presently unclear to what extent it depends on interparticle interactions and thermodynamic conditions. DML may presently only provide a qualitative forecast on the behaviour of PSD in liquids where the model may be suitably applied. This aspect should be deepened in future works, providing quantitative forecasts on "if" and "how" DML predicts the PSD dependence on the nature of intermolecular interactions.

Because the DML provides a way to calculate the relaxation time(s) characterizing the liquid particle - wave packet interaction, an interesting insight is to verify the capability of the model also in predicting the temperature dependence of the relaxation time and to reproduce the Arrhenius or more-than-Arrhenius trend observed in systems exhibiting a structural relaxation. The above questions could find an answer from a Molecular Dynamics Simulation, in which the duality of liquids is considered and the two subsystems interact through the elementary, time reversal interaction described in the paper and at the base of the DML [Peluso , F., et al, in preparation].

The large amount of experiments performed in controlled isothermal conditions [15-16,19-37,61-65,67-72] allowed the discovery of the collective dynamics of liquids and the evaluation of sound velocity at high frequencies. This aspect is interpreted within the DML with the presence in liquids of solid-like dynamic clusters of molecules in continuous rearrangement, in which elastic excitations propagate similarly to phonons as in the corresponding solid phase. The size of these clusters is the phonon mean free path, and they result to be made of 10 to 20 liquid molecules. This vision is supported by the comparison of the relaxation time(s) and of the average lifetime of the phonons. The number density of phonons, of the same OoM as the molecular density, also concurs with the hypothesis of single phonon-*liquid particle* interaction.

Understanding how liquid parameters, as correlations lengths, sound velocity, thermal conductivity, etc, evolve from equilibrium to non equilibrium conditions is an interesting topic that could be achieved by performing experiments in non-stationary temperature gradients. Also very interesting is to understand whether and how a temperature gradient affects the viscous coupling, and this could be understood performing experiments consisting in the application of a temperature gradient to a stabilised isothermal liquid, ensuring that the average temperature remains unchanged, preventing the convection instability and performing light scattering experiments during the transient, exactly the situation that is normally avoided in all the experiments. An alternative way is to heat a small volume of liquid by hitting it with a focalised high power laser beam.

---

accelerations of the liquid would be required (or strong relative velocities of the two surfaces) which would help destroy the thermal phenomenon.

Another type of experiments could be the investigation of the glassy and liquid-to-solid transitions. A technique that could be applied is to start with a liquid medium in a container, with stationary temperature gradient applied to it. By lowering at the same time the temperatures on the two sides of the container, also the average temperature of the medium will be lowered until it solidifies. Light scattering experiments performed during the non stationary phase should allow to investigate the dynamics of the system at the cross-over of the glassy and liquid-to-solid transitions. The local domains shall be oriented following the external temperature gradient; this shall allow in turn the increase of the correlation lengths, of sound velocity, of thermal conductivity, etc., along the preferential direction of the external temperature gradient, the asymmetry introduced into the system. A difference between the values of the parameters when measured along the direction of the temperature gradient and those measured along a different direction should emerge from the collected experimental data. This technique has also the further effect of introducing an anisotropy into the liquid medium, becoming anisotropic the solid phase.

The results obtained by Esposito [98] and Nicolis [118] about the capability of phonons to carry negative mass could be experimentally confirmed by performing two types of experiments; one on ground in a system with an upward temperature gradient to avoid convection; the second in weightlessness, as for instance on orbiting platforms or, even better, in satellites orbiting in the Lagrange point. The latter would allow to measure the net effect of mass carried by phonons without the influence of gravity (in such a system the Clausius Postulate is valid for sure). Having a more clear vision from experiments of how mass and momentum are carried by phonons, may indeed help to understand how to manage their contribution in theoretical studies, and probably answer some of the questions raised in the previous section.

The importance of thermal fluctuations has been recognized since a long time in the dynamics of proteins [see for instance 120-123] and the correlation between the high-frequency dynamics of liquid water and that of protein dynamics should be further investigated. It is also reasonable to expect that gradients of concentrations may give similar results as those due to temperature gradients. Performing such experiments in the study of protein crystal growth might allow to measure the lengths on which aggregation starts and, probably, discover that temperature gradients may help in controlling and addressing correctly the crystal formation.

The long list of references in this paper is by far not complete neither exhaustive, but it represents only a selection of the more relevant contributions to support the topic dealt with. Yet still today there are many questions on the mesoscopic dynamics of liquids that are unanswered, as well as a deficit of experiments. It is intriguing to close this paper by quoting a sentence by Brillouin dating back to 1936 and probably forgotten [4]: "On est donc contraint d'admettre un point de vue intermédiaire: un liquide, à basse température, serait un réseau cristallin défini, mais présentant une ou plusieurs directions de plans de clivage infiniment facile. La distinction, par rapport au liquid idéal, est essentielle: nous aurons, dans de trés petits domaines, des structures cristallines; ces cristaux résistant à un effort tangentiel, si celui-ci s'exerce dans une direction autre que cell de clivage infinement facile." ... "Microscopiquement, (les liquides) sont des solides; macroscopiquement ils apparaitront liquides."

## 7. Dedication and Thanksgiving

The author wants to dedicate this paper *in memoriam* of prof. F.S. Gaeta, who recently passed away. The author wants even to express his gratitude to prof. F.S. Gaeta for his enlightening mentoring.

The author is grateful to prof. A. Romano for having read the manuscript and for the helpful discussions. Finally the author is indebted to the Editor-in Chief prof. P. Ayyaswamy and to the two anonymous reviewers for having improved the quality and readability of the manuscript.

## 8. Figures

*Figure 1*

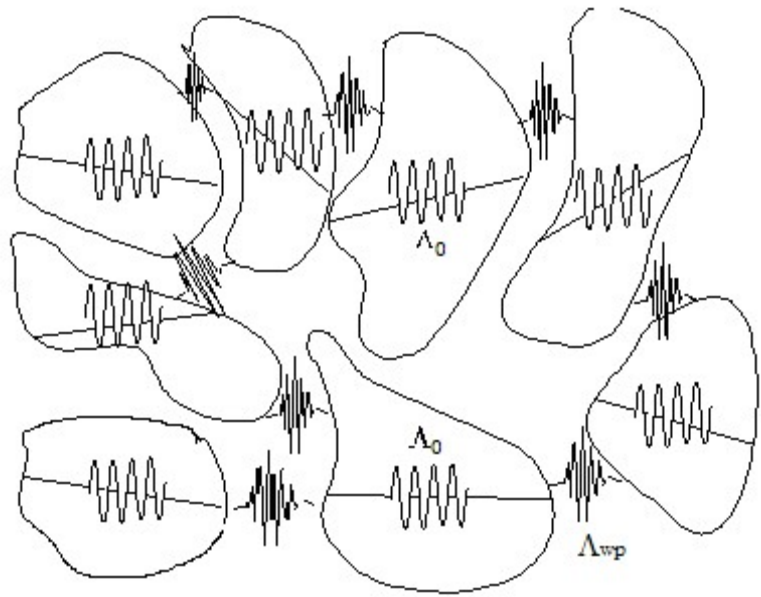

**Figure 1**. Rendering of how ephemeral solid-like structures fluctuate and interact within the amorphous liquid global system at equilibrium.

Modified after Peluso, F., see Reference [76]. MDPI Credits

*Figure 2*

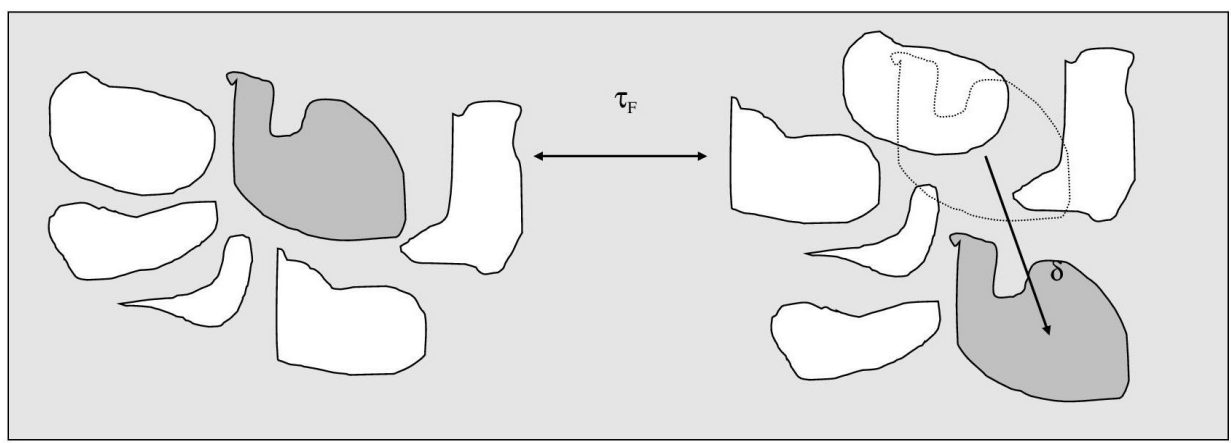

**Figure 2.** A "wandering iceberg" jumping between two equilibrium positions in a liquid. The jump takes place after an average relaxation time lasting $\tau_F$, during which the liquid particle moves by a distance $\delta$, it can be associated to Nonaffine displacements working in disordered systems like liquids.

*Figure 3*

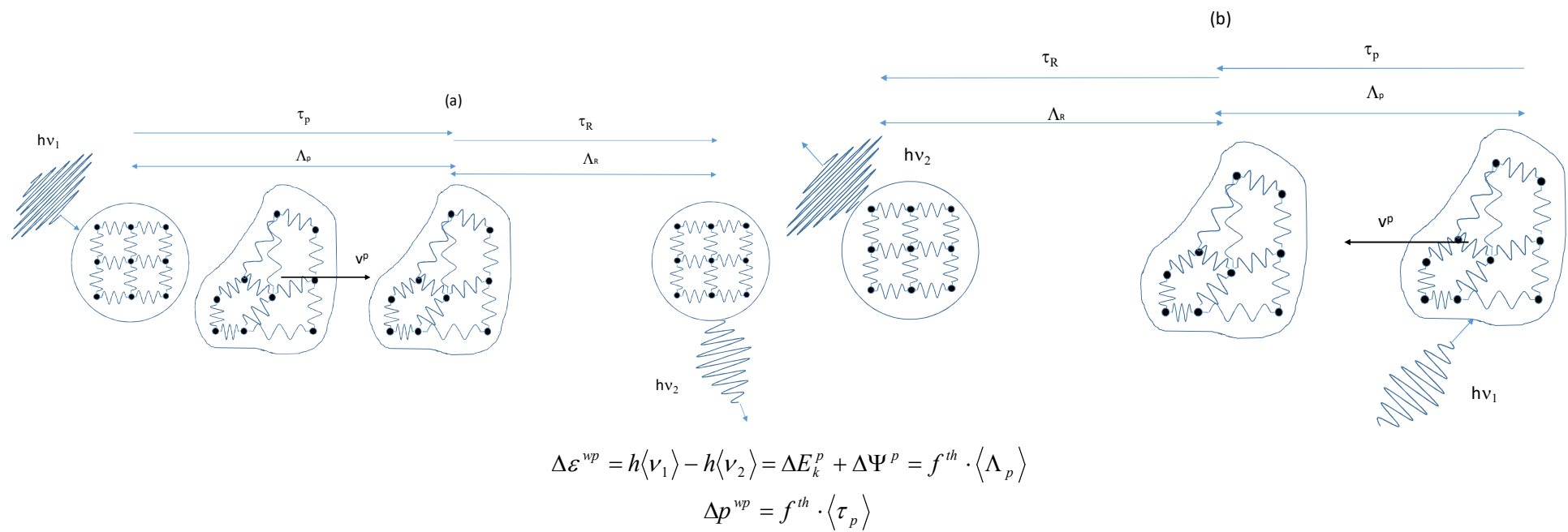

**Figure 3**. Schematic representation of inelastic collisions of wave-packets with *liquid particles*.

*Figure 4*

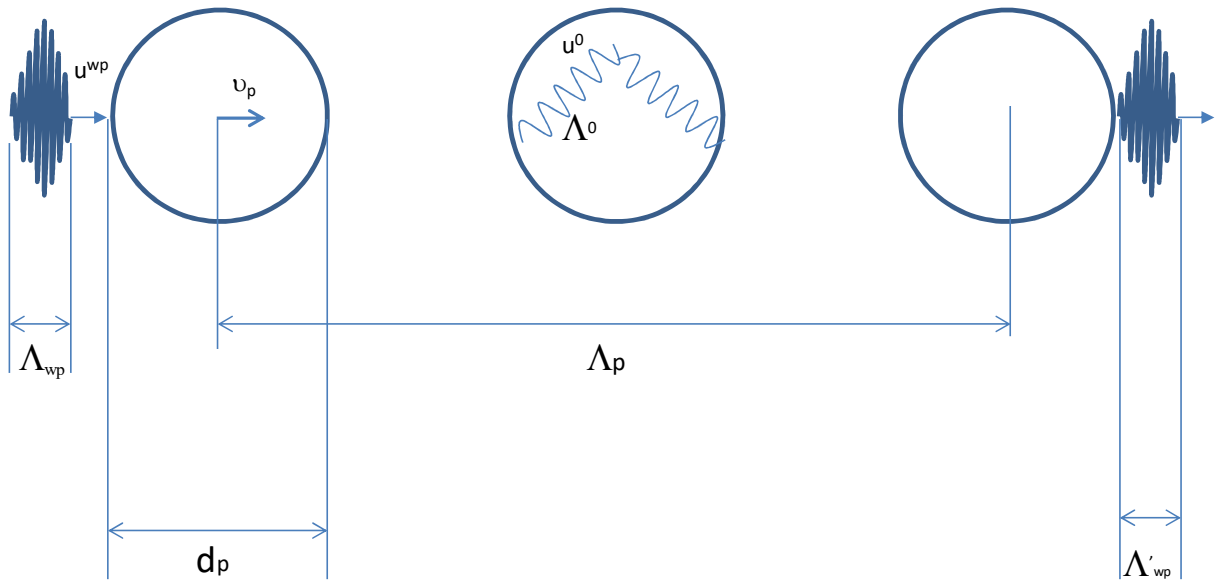

$$\langle \Delta^{ph} \rangle = \langle \Lambda_{wp} \rangle + \langle d_p \rangle + \langle \Lambda_p \rangle + \langle \Lambda'_{wp} \rangle$$

$$\langle \tau^{ph} \rangle = \frac{\langle \Lambda_{wp} \rangle}{u^{wp}} + \frac{d_p}{u^0} + \frac{\langle \Lambda_p \rangle}{u^0} + \frac{\langle \Lambda'_{wp} \rangle}{u^{wp}}$$

**Figure 4.** Close-up of the wave-packet – particle interaction shown in Figure 3a. Only the first part of the interaction is represented, i.e. that during which the wave packet transfers momentum and energy to the *liquid particle*.

Modified after Peluso, F., see Reference [76]. MDPI Credits

*Figure 5*

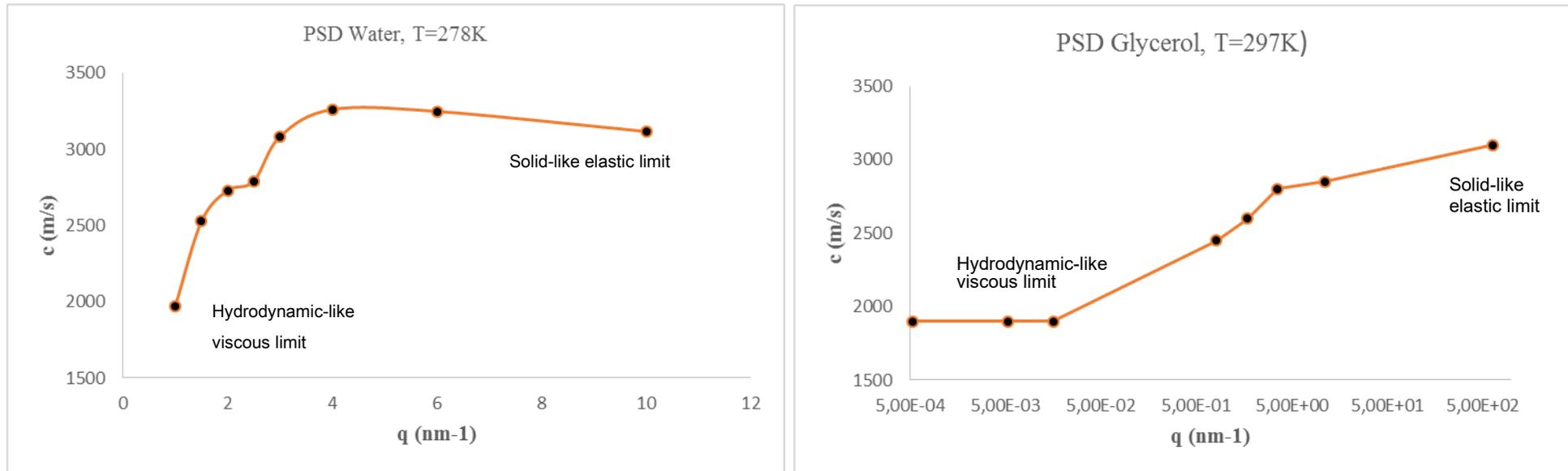

**Figure 5.** Positive Sound Dispersion – PSD – for Water and Glycerol at the indicated temperature and momentum values. Data are from [34] for Water and from [19] for Glycerol.

# 9. Tables

*Table 1*

| Parameters | Variables | Eqs.(#) |
|---|---|---|
| $\eta_l = 9 \cdot 10^{-4}\, Pa \cdot s$ | $35 \leq \upsilon_p \leq 176\ m/s$ | 31 |
| $\nu^{wp} = 0.9;\ 1.0;\ 1.5;\ 2.0\ ;3.0 \cdot 10^{12}\ Hz$ | $0.29 \leq \Lambda_p \leq 1.20 \cdot 10^{-9}\ m$ | 38 |
| $\alpha = 0.3;\ 0.5;\ 0.7$ | $4.0 \leq \tau_p \leq 13.0 \cdot 10^{-12}\ s$ | 34 |
| $n = 5$ | $8.9 \leq \Delta^{ph} \leq 22.9_6 \cdot 10^{-9}\ m$ | 32 |
| $m_p = 3;5;10\, m_{H_2O} = 9;\ 15;\ 30 \cdot 10^{-26}\ kg$ | $1.3 \leq \tau_\eta \leq 8.8 \cdot 10^{-15}\ s$ | 42 |
| $r_p = 1.0,\ 1.5,\ 2.0 \cdot 10^{-9}\ m$ | $0.045 \leq \Lambda_\eta \leq 1.5 \cdot 10^{-12}\ m$ | 43 |

**Table 1.** List of parameters (1st column) introduced to evaluate the several variables (2nd column) characterizing the mesoscopic dynamics of liquids, and of the corresponding values for the named parameters. In the 3rd column the number of equation is listed that has been used to calculate the variables listed in the 2nd column.